\documentclass[11pt]{article}
\usepackage{amssymb, amsmath, amsfonts, amsthm} 
\usepackage{graphicx}

\newcommand{\Eq}[1]     {(\ref{#1})}

\newcommand{\Thm}[1]     {Thm.~\ref{#1}}
\newcommand{\Lem}[1]    {Lem.~\ref{#1}}

\newcommand{\Fig}[1]    {Fig.~\ref{#1}}
\newcommand{\Sec}[1]    {\S\ref{#1}}

\newcommand{\sgn}{\mathop{\rm sgn}}
\renewcommand{\mod}{\mbox{mod }}
\newcommand{\R}{{\mathbb R}}
\newcommand{\Z}{{\mathbb Z}}
\newcommand{\N}{{\mathbb N}}
\newcommand{\T}{{\mathbb T}}
\newcommand{\C}{{\mathbb C}}
\newcommand{\bS}{{\mathbb S}}

\newcommand{\calS}{\mathcal{S}}
\newcommand{\calT}{\mathcal{T}}
\newcommand{\calU}{\mathcal{U}}

\newcommand{\om}{\omega}

\newtheorem{theorem}{Theorem}
\newtheorem{lem}[theorem]{Lemma}

\theoremstyle{definition}

\theoremstyle{definition}

\newcommand{\InsertFig}[4]{
\begin{figure}[thbp]
  \centerline{
    \includegraphics[width=#4]{#1.eps}
  }
  \caption{{\footnotesize  #2.eps}}
  \label{#3}
\end{figure}}

\newcommand{\InsertFigFour}[7] {
\begin{figure}[htbp]
  \centerline{
    \includegraphics[width=#7]{#1.eps}
    \hskip 0.02in
         \includegraphics[width=#7]{#2.eps}
       }
  \centerline{
    \includegraphics[width=#7]{#3.eps}
    \hskip 0.02in
         \includegraphics[width=#7]{#4.eps}
       }
       \caption{{\footnotesize  #5.eps}}
       \label{#6}
\end{figure}}

\begin{document}
\title{Numerical Computation of the Stable and Unstable Manifolds of Invariant Tori}
\author{Derin B. Wysham and James D. Meiss\thanks{JDM acknowledges support from from NSF grant
DMS-0202032 and DBW would like to thank VIGRE grant number DMS-9810751. We would like to thank
Hector Lomel\'i for the original suggestion of this method, and Angel Jorba for useful discussions.}\\
		Univ. of Colorado, Boulder\\
		Applied Mathematics\\
		Boulder  CO  80309-0526}
\maketitle
\begin{abstract} We develop an iterative
		technique for computing the unstable and stable eigenfunctions of the invariant tori of
		diffeomorphisms.  Using the approach of Jorba, the linearized equations are rewritten
		as a generalized eigenvalue problem.  Casting the system in this light allows us to
		take advantage of the speed of eigenvalue solvers and create an efficient method
		for finding the first order approximations to the invariant manifolds of the torus.
		We present a numerical scheme based on the power method that
		can be used to determine the behavior normal to such tori, and give some examples of
		the application of the method. We confirm the qualitative conclusions  of the 
		Melnikov calculations of Lomel\'i and Meiss (2003) for a volume-preserving mapping.
\end{abstract}

\section{Introduction}
\textbf{Hyperbolic sets commonly occur in dynamical systems; consequently, developing an understanding of the structure of their invariant manifolds is of considerable significance. For example, the stable and unstable invariant manifolds of a hyperbolic fixed point often intersect transversally, and the resulting heteroclinic tangle is a primary mechanism for the onset of chaos \cite{Guckenheimer83,Wiggins94}. The theory of transport in dynamical systems is based on the construction of 
regions bounded by partial barriers that can often be built using segments of stable and 
unstable manifolds \cite{Meiss92, Wiggins92}. Moreover, Arnold's mechanism for drift 
in near integrable Hamiltonian systems is also based on the construction of 
transition chains made up of invariant manifolds \cite{Arnold64,Lochak99}. In this paper we will develop a simple iterative method for the computation of such manifolds that was suggested to us by Hector Lomel\'i.}

The easiest algorithms for computing invariant manifolds of a hyperbolic set for a discrete dynamical system $f$ are based on iteration. For example, suppose that $f$ has a hyperbolic fixed point $p$. Near $p$, $W^u(p)$ can be approximated by the linearized unstable space, $E^u(p)$, or more accurately as the fixed point of a contraction map on sequences of points in a neighborhood of $p$ \cite{Homburg03}. Given such an approximation, one can start the iteration with a fundamental domain $\calU$ on $W^u(p)$, i.e. a shell enclosing $p$ whose inner boundary iterates to its outer boundary \cite{Lomeli00a}. Iteration of $\calU$ will then cover $W^u(p)$. The shell $\calU$ can be covered with a grid of points on several spheres. Since distances between grid points typically grow exponentially with iteration, it is worthwhile to use distance as a diagnostic for inserting or removing additional grid points as needed. This method works well for one-dimensional manifolds \cite{Hobson93}, and for two-dimensional manifolds when the two multipliers have the same magnitude \cite{Lomeli98, Lomeli00a}. More sophisticated methods for constructing invariant manifolds define a flow on the manifold that does not suffer from the problems of exponential nonuniformities when the multipliers have distinct magnitudes \cite{Krauskopf98a,Krauskopf98b,Guckenheimer04}.

More generally, let $f:{\R}^n \rightarrow {\R}^n$ be a diffeomorphism that defines the discrete dynamical system
\begin{equation}\label{eq:map}
	y'=f(y) \;, 
\end{equation}
and suppose that $f$ has a compact invariant set $\calT$. Given a point $z \in \calT$, the linearization of
the dynamics near $\calT$ defines a skew-product system on
$(x,z) \in \R^n \times \calT$
\begin{align}\label{eq:cocycle}
    x_{t+1} &= Df(z_t)x_t \;,\nonumber\\
    z_{t+1} &= f(z_t) \;.
\end{align}
This is an example of a cocycle over the dynamical system $f|_{\calT}$ \cite[Supplement]{Katok99}. Many properties of such systems are well-known.

We will specialize to the case when $\calT$ is an embedded invariant torus. Invariant tori commonly occur in dynamical systems. For example, KAM theory implies that nearly integrable, symplectic twist maps have Cantor sets of tori. Tori whose dimensions are less than the number of degrees of freedom of the system can be normally hyperbolic \cite{Treshev94, Bolotin00}. Hyperbolic, or ``whiskered" tori provide a primary mechanism for transport in these systems and are a necessary part of Arnold's construction of a transition chain for systems with more than two degrees of freedom. Invariant tori are also a common feature of volume-preserving mappings \cite{Piro88,Feingold89,Huang98,Mezic98,Cartwright02, Lomeli00a,Lomeli03}.  Our goal in this paper is to develop an iterative algorithm for constructing the linear stable and unstable spaces of an invariant torus. 

Suppose that the function $z: {\T}^r \rightarrow \calT \subset {\R}^n$ represents the embedding of the invariant $r$-dimensional torus $\calT$. We will use angle coordinates $\theta$ on $\T^r$; they are taken $\mod 1$, so that $z(\theta+m) = z(\theta)$ for all $m \in \Z^r$. We assume that the dynamics on $\calT$ is conjugate to the rigid rotation $\theta \rightarrow \theta + \om$ with rotation vector $\om \in \R^r$ so that
\begin{equation}\label{eq:conjugacy}
	f(z(\theta))=z(\theta+\om) \;.
\end{equation}
For this case the cocycle \Eq{eq:cocycle} can be written
\begin{align}\label{eq:skewprod}
      x'    &=A(\theta)x \;, \nonumber \\
     \theta'&=\theta+\om \;,
\end{align}
where $A: \T^r \rightarrow GL(n)$ is defined by
$A(\theta) \equiv Df(z(\theta))$.

We will always assume that $\om$ is {\em incommensurate}:
\begin{equation}\label{eq:incommensurate}
	  \om \cdot m \neq n\quad \forall \;\; m \in \Z^r \setminus \{ 0\},\;n\in \Z \;.
\end{equation}
Denoting the distance to the nearest integer lattice point as 
\begin{equation}\label{eq:pseudonorm}
         ||\alpha||_{\Z^k} \equiv \min_{n\in\Z^k} || \alpha - n|| \;,
\end{equation}
for an $\alpha \in \R^k$,\footnote
{
  Note that this expression  is not a norm.
} 
then a rotation vector $\om$ is incommensurate if $||m \cdot \om||_\Z$ is nonzero whenever $m$ is nonzero. When $\om$ is incommensurate, the system \Eq{eq:skewprod} is a 
\emph{linear, quasiperiodic skew-product}. 

Often it will also be necessary to assume that the rotation vector $\om$ satisfies a {\em Diophantine condition}, i.e., that there exist $c >0$ and $\tau \ge r$ such that
\begin{equation}\label{eq:diophantine}
         ||m \cdot \om||_{\Z} > c/||m||^\tau, \, \forall m \in \Z^r \setminus \{ 0\}\;.
\end{equation}
For example, KAM tori are Diophantine. As we will see next, many of the theorems that show \Eq{eq:skewprod} can be simplified, or ``reduced," rely on such conditions as well.
We will often use the golden mean
\begin{equation}\label{eq:goldenMean}
   \gamma \equiv \frac{1+\sqrt{5}}{2} \;,
\end{equation}
which is Diophantine and has the maximal value for $c$.
\section{Reducibility} 

Following Lyapunov,  a skew-product is \emph{reducible} if there is a continuous change of variables $C: \T^r \rightarrow GL(n)$, so that defining $x=C(\theta)y$, \Eq{eq:skewprod} becomes
\begin{align}\label{eq:reduced}
      y'    &=Jy \nonumber\\ 
     \theta'&=\theta+\om \;,
\end{align}
such that the matrix
\begin{equation}\label{eq:Jdefine}
     J\equiv (C(\theta+\om))^{-1}A(\theta)C(\theta)
\end{equation}
does not depend on $\theta$ \cite{Eliasson98,Jorba01}. The reducing transformation is the quasiperiodic generalization of the Floquet representation in the periodic case. Just like in the Floquet case, the transformation $C$ and the matrix $J$ are generally complex; however, they can be made real at the expense of allowing $C$ to act on some finite cover of $\T^r$ \cite{Johnson81, Ellis82}.

The main point of reducing the matrix is that 
the fundamental matrix solution of \Eq{eq:skewprod},
\begin{equation}\label{eq:Andef}
	A^{(t)}(\theta) \equiv \prod_{j=0}^{t-1} A(\theta+j\om) \;,
\end{equation}
then has the representation
\[
      A^{(t)}(\theta) = C(\theta+t\om) J^t (C(\theta))^{-1} \;.
\]
Thus the stable, unstable, and center spaces of \Eq{eq:skewprod} are easily found from those of the constant matrix $J$. Since the dynamics are now are
trivial, the case that \Eq{eq:skewprod} is reducible is ideal.  However in contrast with the Floquet case, a reducing transformation cannot always be found.
When $A$ is smooth and close enough to a parameter-dependent, constant matrix and $\om$ is Diophantine \Eq{eq:diophantine}, then $A$ is reducible for all but a small, positive-measured set of parameter values \cite{Jorba92}.  On the other hand, some systems with strongly varying $A$, like the skew product version of the quasiperiodic Schr\"{o}dinger equation
\begin{equation*}
\epsilon(u_{n+1}+u_{n-1})+V(\theta)u_n=Eu_n
\end{equation*}
with $|\epsilon|<\epsilon_0(V,\om)$ are not necessarily reducible \cite{Eliasson01, SinaiChul91, Frohlich90, Eliasson97}.

One way to try to find a reducing transformation $C(\theta)$ is to find solutions to the eigenvalue problem \cite{Jorba01}
\begin{equation}\label{eq:eproblem}
	A(\theta) \psi (\theta)=\lambda T_{\om} \psi(\theta),
\end{equation}
where $T_\om$ is the translation operator, $[T_\om \psi](\theta) \equiv \psi(\theta+\om)$.
The goal is to find a set of eigenfunctions $\psi \in C({\T}^r,\;{\C}^n) \setminus \{ 0 \}$ and their corresponding eigenvalues $\lambda \in {\C}$. It is easy to see that if $(\lambda, \psi(\theta))$ is an eigenvalue-eigenfunction pair for
\Eq{eq:eproblem},  then for any $m \in \Z^r$,  
\[
	(\lambda e^{2 \pi i m\cdot \om}, \psi(\theta) e^{-2 \pi i m \cdot \theta})
\]
is as well. A pair of eigenvalues such that $\lambda_2 = \lambda_1 e^{2 \pi i m\cdot\om}$ are called ``$\om$-related."

Whenever \Eq{eq:skewprod} is reducible then eigenvectors for \Eq{eq:eproblem} exist; indeed, it is not hard to prove the following result.     


\begin{lem}[\cite{Jorba01}]\label{thm:spectrum}
	Suppose $A \in C(\T^r,GL(n))$ is reducible to the constant matrix 
	$J$ through \Eq{eq:Jdefine}. Then
		\begin{itemize}
		\item if $J$ has an eigenvalue-eigenvector pair $(\lambda, v)$ of $J$, the eigenvalue problem
		\Eq{eq:eproblem} has a solution $(\lambda, C(\theta)v)$, and
		\item if $\lambda$ is an eigenvalue of \Eq{eq:eproblem}, then $\exists m \in {\Z}^r$ such that
		$\lambda e^{2 \pi i m \cdot\om}$ is an eigenvalue of $J$.
		\end{itemize}
	Conversely, suppose there exist eigenfunctions 
	$\psi_j(\theta) \in C(\T^r,\C^n), \; j = 1,\ldots, n$ for 
	\Eq{eq:eproblem} that are independent at each value of $\theta$. Then $A$ is 
	reducible  to the constant matrix $J = \mbox{diag}(\lambda_1, \lambda_2, \ldots,\lambda_n)$ 
	with $C(\theta) = [\psi_1,\psi_2,\ldots,\psi_n]$.
\end{lem} 

\noindent
Note that $\om$-related eigenfunctions are linearly dependent for each $\theta$, even though as functions on the torus they are independent over $\C^n$. Since the set $\{\exp(-i 2 \pi k \cdot \theta): k \in \Z^r\}$ is a Fourier basis for C($\T^r, \C)$, when there is a set of eigenfunctions $\psi_j(\theta)$ that are independent for each $\theta$, then the set of functions
\[
      \{e^{-i 2 \pi k\cdot \theta} \psi_j(\theta): j = 1, \ldots, n, k \in \Z^r\}
\]
is a basis for $C(\T^r,\C^n)$.

In addition to the right eigenfunctions, \Eq{eq:eproblem}, we will use left eigenfunctions,
$\phi \in C(\T^r, \C^n) \setminus \{0\}$, which are defined to be solutions of  
\begin{equation}\label{eq:left}
	A^*(\theta) T_{\om}\phi (\theta)=\lambda \phi (\theta)\;,
\end{equation}
where $A^* = \bar{A}^T$ is the Hermitian conjugate of $A$. The left and right eigenfunctions (if they exist) are orthogonal for each $\theta$. To prove this, we first state a useful lemma (see, e.g. \cite{Los89}).

\begin{lem}\label{lem:toruseq}
	    Suppose $g \in C(\T^r, \C)$, and 
	\begin{equation}\label{eq:toruseq}
	        g(\theta) = a g(\theta + \om)
	\end{equation}
	for incommensurate $\om \in \R^r$ and some constant $a \in \C$. Then either $g \equiv 0$ or there 
	is an $m \in \Z^r$ and a constant $c$ such that $a = e^{-2\pi i m \cdot \om}$ and $g = c e^{2\pi i m\cdot \theta}$.
\end{lem}

\begin{proof}
Expand $g$ in a Fourier series on the torus with coefficients $g_m$, $m \in \Z^r$. If $g \not\equiv 0$ then there is least one $m$ such that $g_m \neq 0$. The Fourier transform of \Eq{eq:toruseq} implies that $g_m = a e^{2\pi i m \cdot \om} g_m$, so that $a =e^{-2\pi i m \cdot \om}$. Suppose there is some other nonzero Fourier coefficient, say $g_l$, then $ae^{2\pi i l \cdot \om} = 1$, which implies $(l-m)\cdot \om \in \Z$, but  by \Eq{eq:incommensurate} the only solution of this is $l = m$. Thus $g$ has a single nonzero Fourier coefficient.
\end{proof}

\begin{lem}\label{lem:orthogonal}
	If $\om$ is  incommensurate, the right \Eq{eq:eproblem} and left \Eq{eq:left} eigenfunctions corresponding to $\om$-unrelated eigenvalues are orthogonal for each $\theta$.
\end{lem}
\begin{proof}
Multiply $\phi_j^*(\theta+\om)A(\theta)=\lambda_j\phi_j^*(\theta)$ by $\psi_k(\theta)$ on the right,  
and multiply $A(\theta)\psi_k(\theta)=\lambda_k\psi_k(\theta+\om)$ by  $\phi_j^*(\theta+\om)$ on the left.  Then subtract the results to obtain
\begin{eqnarray*}
	\lambda_j \phi_j^*(\theta)\psi_k(\theta)=\lambda_k\phi_j^*(\theta+\om)\psi_k(\theta+\om).
\end{eqnarray*}
If $j=k$, then this equation implies that  $\phi_k^*(\theta)\psi_k(\theta)$ is constant on a dense set (since $\om$ is incommensurate) and so must be constant everywhere by continuity. The constant can be chosen to be one without loss of generality. If $j \neq k$, then  \Lem{lem:toruseq} implies either that $\lambda_j = \lambda_k e^{-2\pi i m \cdot \om}$, which would mean that the two eigenvalues are
$\om$-related, or $\phi_j^*(\theta)\psi_k(\theta) \equiv 0$.
\end{proof}

Just as a constant matrix need not be diagonalizable, a complete set of eigenfunctions of $A$ need not exist, even when $A$ is reducible. To deal with this case, define a {\em generalized eigenfunction} for eigenvalue $\lambda$ to be a nonzero, continuous solution to
\begin{equation}\label{eq:geneigen}
	\prod_{t=0}^{k-1}\left[ A(\theta+t\om)-\lambda T_{\om}I\right] \psi(\theta)=0 \;,
\end{equation}
for some $k \in \N$. Suppose that for a fixed $\lambda$, there exist exactly $l$ continuous solutions $\psi_j, j = 1,\ldots,l$  to \Eq{eq:geneigen} that are independent for each $\theta$. These are a basis for an $l$-dimensional vector space $E_{\lambda} \subset C(\T^r,\C^n)$. It is easy to see that $ E_\lambda$ is an invariant subspace under $A$. Indeed, if $\psi \in E_{\lambda}$, then
\[
  \left(A(\theta-\om) - \lambda T_\om I \right)\psi(\theta-\om) 
\]
is in $E_\lambda$ as well since it is annihilated by \Eq{eq:geneigen}. 

Just as there is a correspondence between regular eigenfunctions of $A$ and eigenvectors of $J$, there is also a correspondence between generalized eigenfunctions and the generalized eigenvectors of $J$.

\begin{lem}\label{lem:geneigen}
    The matrix $A(\theta)$ is reducible if and only if there exists a set of generalized eigenfunctions $\psi_j \in C(\T^r,\C^n), \, j = 1,\ldots, n$ that are linearly independent for each $\theta$. 
\end{lem}

\begin{proof}
Suppose that $A$ is reducible to $J$ by the transformation $C(\theta)$. If $w$ is a generalized
eigenvector of $J$ with eigenvalue $\lambda$, then for some $k \in \N$, $\left( J-\lambda I\right)^kw =0$.
Defining $\psi(\theta) = C(\theta) w$, 
\begin{align*} 
	\left( J-\lambda I\right)w 
	       &= (C(\theta+\om))^{-1}C(\theta+\om)\left( J-\lambda I\right)C^{-1}(\theta)C(\theta)w\\
	       &= (C(\theta+\om))^{-1}\left[A(\theta) -\lambda T_{\om} I \right]\psi(\theta) \;. 
\end{align*}
Applying $(J-\lambda I)$ an additional $k-1$ times shows that $\psi$ satisfies \Eq{eq:geneigen}. Since $C(\theta)$ is invertible, linear independence of the generalized eigenvectors implies the linear independence of the corresponding $\psi_j$.

To prove the converse, let $E = \mbox{span}(\psi_1,\psi_2,\ldots,\psi_n) \subset C(\T^{r},\C^{n})$. Since
$E$ is a finite-dimensional vector space, the standard representation result for generalized eigenfunctions \cite[Appendix III] {Hirsch74} implies that $E$ is the direct sum of generalized eigenspaces $E_{\lambda_j} , \; j = 1, \ldots, l$, corresponding to eigenvalues $\lambda_j$, and that each of these has a basis $\hat{\psi}^k_j \;, k = 1,\ldots n_j = \dim E_{\lambda_j}$
such that
\begin{align*}
    \left[A(\theta) - \lambda_j T_\om\right] \hat{\psi}^{1}_j (\theta) &= 0 \;, \\
    \left[A(\theta) - \lambda_j T_\om\right] \hat{\psi}^{k+1}_j (\theta) &= \hat{\psi}^k_j(\theta+\om)
       \;,\;  k = 1 \ldots n_j-1 \;.
\end{align*}
Define the matrix $C(\theta) = [\hat{\psi}^1_1(\theta), \dots,\hat{\psi}^{n_l}_l(\theta)]$. By assumption, 
$\det C(\theta) \not \neq 0$.
Then by the reverse of the previous argument, it is easy to see that  
$J(\theta)=(C(\theta+\om))^{-1}A(\theta) C(\theta)$ does not depend on $\theta$, and has 
generalized eigenvectors $w_i = (C(\theta))^{-1}\psi_i(\theta)$.
\end{proof}

In this paper, the skew-product system \Eq{eq:skewprod} is the linearization of a map on an embedded invariant torus. For this case there are always unit eigenvalues: 
\begin{lem} \label{lem:tangent}
	If $f$ has an embedded $r$-dimensional invariant torus on which the motion is conjugate
	to a rigid rotation, then the eigenvalue 
	problem \Eq{eq:eproblem} associated with the linearization \Eq{eq:cocycle} 
	has $r$ linearly-independent eigenfunctions $\psi_{i}(\theta)=\frac{\partial z}
	{\partial \theta_i}(\theta),\;i=1,\dots,r$ with $\lambda = 1$.
\end{lem}
\begin{proof}
Differentiation of the conjugacy \Eq{eq:conjugacy} with respect to $\theta_i$ shows that the
functions $\psi_i$ are eigenfunctions with eigenvalue one.  Since $z(\theta)$ is an embedding,
$D_{\theta}z$ has rank $r$ for all $\theta$ in ${\T}^r$, so the eigenfunctions are independent.
\end{proof}

\noindent
The reducibility of $A$ thus hinges on the existence of $n-r$ generalized eigenfunctions that are transverse to the tangent space of the invariant torus.  When $\om$ is incommensurate, each eigenvalue $\lambda$ is associated with a dense set of $\om$-related eigenvalues. Thus the closure of the spectrum of \Eq{eq:eproblem}
will correspond to the union of up to $n-r+1$ circles centered at the origin of the complex plane.  As an illustration, \Fig{fig:fepsevals} shows the an approximate spectrum for a three-dimensional mapping (this mapping is studied in \Sec{sec:examples}) with an invariant circle. For this figure, the eigenvalues were computed using a Fourier series algorithm developed by Jorba \cite{Jorba01}; only finitely many eigenvalues are found due to truncation of the Fourier series.
\InsertFig{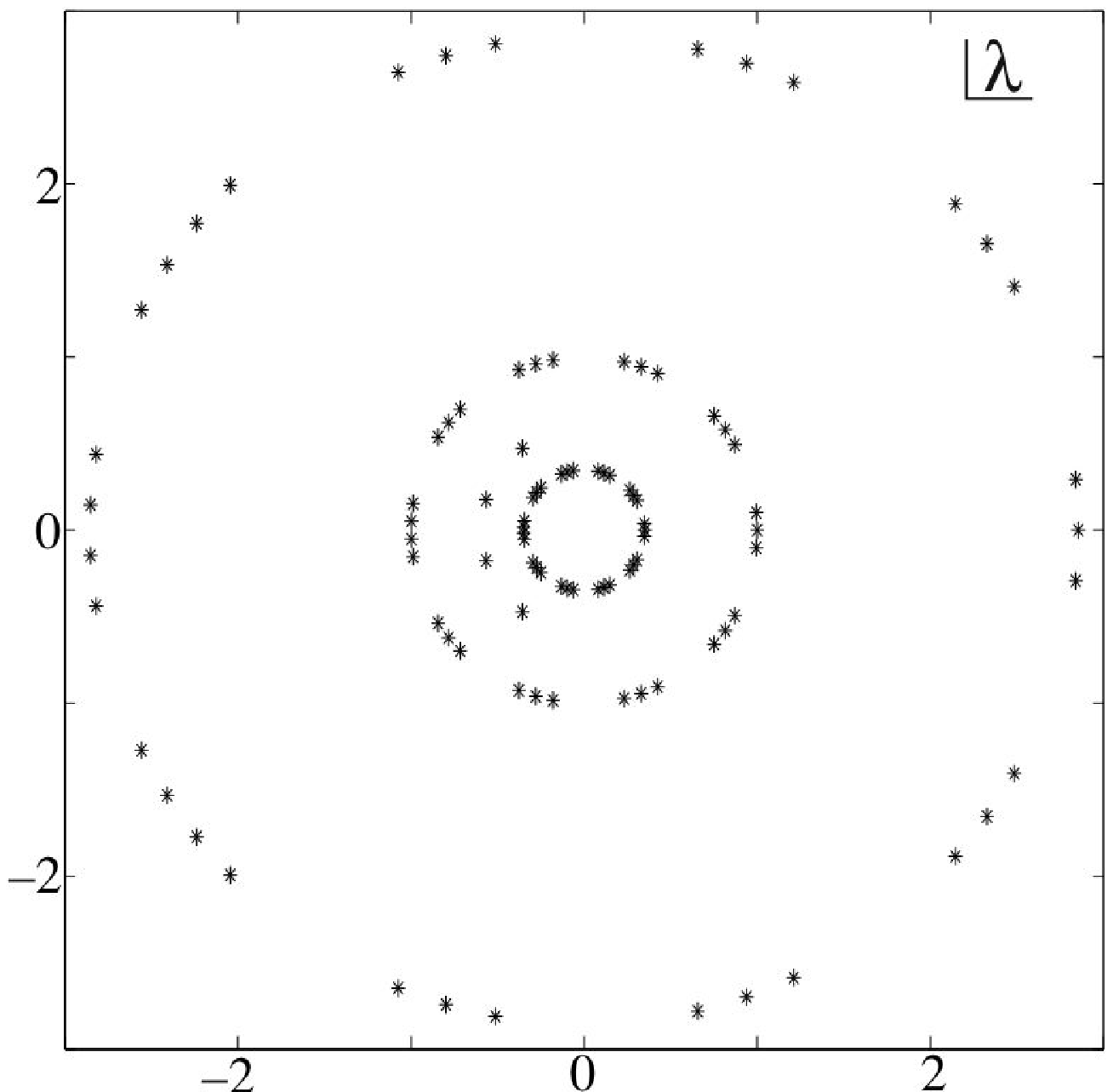}
	{Numerically generated spectrum of \Eq{eq:eproblem} for the volume-preserving example 
	$f_{\epsilon}$ \Eq{eq:lomeliMap} using $16$ complex Fourier modes. Here the parameters are
	$\epsilon=0$, $\nu=.35$, and $\om=\gamma^{-1}$, the inverse of \Eq{eq:goldenMean}.
	There are $99 = 3(2\times 16 +1)$ eigenvalues shown; all but four fall very nearly one 
	of three circles whose radii correspond to the three eigenvalues $\lambda = \nu, 1,\nu^{-1}$. The final four
	values are are numerical artifacts.}
	{fig:fepsevals}{3.5 in}

The eigenvalues of \Eq{eq:eproblem} are related to the Lyapunov multipliers of the skew-product \Eq{eq:skewprod}.  According to Oseledec's theorem, $\R^n$ can be decomposed into invariant subspaces spaces $H_i(\theta),\; i = 1\ldots k(\theta)$ for almost all $\theta$ such that for each $x_0 \in H_i(\theta)$, the Lyapunov multiplier,
\[
    \mu_i(x_0) = \lim_{t\rightarrow \infty} ||A^{(t)}(\theta) x_0||^{1/t} \;,
\]
exists. Moreover the matrix
\[
	\Lambda(\theta)\equiv \lim_{t\rightarrow\infty}
	\left[
	    \left(A^{(t)}(\theta)\right)^*A^{(t)}(\theta)
	\right]^{1/2t} \;
\]
exists for almost all $\theta$. An eigenvector $v(\theta)$ of $\Lambda$ is called a Lyapunov vector   \cite{jalnine04}. The corresponding Lyapunov multiplier $\mu$ is the growth rate of $v$ under iteration
\begin{equation}\label{eq:lyapunov}
	\mu(v) = \lim_{t\rightarrow\infty}\|A^{(t)}(\theta)v(\theta)\|^{1/t} \;.
\end{equation}

In the ``nicest" case, $\calT$ is uniformly hyperbolic, and then the spectral subspaces $H_i(z)$ vary continuously and the corresponding the Lyapunov multipliers are constant and exist everywhere. When $A$ is reducible, the limits exist uniformly for all $\theta$ so that the matrix $\Lambda$ is continuous. In this case the Lyapunov spectrum coincides with the magnitudes of the eigenvalues $\lambda_i$. A slightly weaker converse is also true:

\begin{theorem}[\cite{Eliasson98,Johnson81}]\label{thm:lyapunov}
	If $A(\theta)$ has a uniform and simple Lyapunov spectrum $\mu_1>\mu_2>\ldots\mu_n$ 
	and if $\omega$ satisfies a Diophantine condition \Eq{eq:diophantine}, then $A$ is reducible.
\end{theorem}

\noindent
We will use this result in the next section to compute the eigenfunctions.

The results in this section reveal the connections between the reducibility of a skew product and
generalized eigenvectors: the transformation that reduces the linear quasiperiodic skew-product \Eq{eq:skewprod} is constructed from $n$ linearly independent generalized eigenfunctions \Eq{eq:geneigen} and their corresponding $\om$-unrelated eigenvalues. These functions will form the transformation matrix $C(\theta)$.

Our goal is to use the eigenfunctions as the linear approximations to the invariant manifolds $W^c(z(\theta))$, $W^u(z(\theta))$, and $W^s(z(\theta))$.

\section{Computing Eigenfunctions}

One method for finding the eigenfunctions of \Eq{eq:eproblem} is to expand $T_{\om},A(\theta)$ and $\psi(\theta)$ into Fourier series and use the fact that the Fourier coefficients of a product are convolutions of the coefficients of the individual functions. If the series are truncated, then the resulting equations correspond to a finite dimensional linear system, and they can be solved by standard numerical techniques \cite{Jorba01}. Iterative refinement methods have also been developed that use Fourier series \cite{Haro04}.
This method has several drawbacks.  First of all, the construction of the representation of \Eq{eq:eproblem} in a real Fourier basis is somewhat involved. Moreover, the dimensionality of the resulting operators $A(\theta)$ and $T_{\om}$ can be extremely large, especially if $r$ is large. A representation with $N$ complex Fourier modes and their conjugates on $\T^r$ requires $(2N+1)^r$ coefficients. Since each component of $\psi$ requires such a representation, the truncated representation will be of length $n (2N+1)^r$.  In our experience, a minimum of $N=32$ modes is required for reasonable (twelve digit) accuracy even in very smooth examples.\footnote
{
  For example,  $f_{\epsilon}$ in \Eq{eq:perturbedLomeli} with $\epsilon<0.35$.
}
Thus, if $n=3$, and the invariant set is a circle, the representation of $A$ requires a matrix of size $195 \times 195$. If the space were five-dimensional and the invariant set where a three torus, the matrices would have a minimum size of the order of $10^6\times 10^6$.  Even the most reliable eigenvector solvers could prove unwieldy in such a situation. 

Thus we are motivated to develop an iterative method for finding eigenfunctions.
\subsection{Power Method}\label{sec:pwrmeth}

When the eigenvalues of a constant matrix have distinct magnitudes, the power method can be an effective technique for finding the eigenvectors. We will use this idea here to find the eigenfunctions of $A(\theta)$ transverse to the torus (recall that the remaining $r$ tangent eigenfunctions are known by \Lem{lem:tangent}).  While the power method will find the dominant eigenfunction as long as $A$ is reducible and the dominant Lyapunov multiplier is distinct, for the remainder of this section we will assume that $A(\theta)$ is a real matrix with a uniform and simple transverse Lyapunov spectrum. Thus the transverse Lyapunov multipliers have distinct magnitudes:
\[
	\mu_1 > \mu_2\ >...\ldots > \mu_{n-r}, \quad 
	\mu_i \neq 1, \; i = 1 \ldots n-r \;.
\]
If we also assume that $\om$ satisfies a Diophantine condition \Eq{eq:diophantine}, then \Thm{thm:lyapunov} applies and $A$ is reducible. Since the multipliers are distinct and $A$ is assumed to be real, there will be a corresponding set of real eigenvalues $\lambda_i = \pm \mu_i$ and real eigenfunctions $\psi_i$ of \Eq{eq:eproblem}. Algorithms for the more general case of multiple and complex eigenvalues will be discussed in a future paper \cite{Wysham05}.

As in the power method, we begin with an arbitrary initial vector $q^{(0)}$.  Since $A$ is assumed reducible, every vector can written as a linear combination of the eigenfunctions at some $\theta_0$:
\begin{equation}\label{eq:initialguess}
	q^{(0)}=\sum_{i=1}^{n}\alpha_i\psi_i(\theta_0),
\end{equation}
and generically $\alpha_1\neq 0$. Iteratively define sequences $u^{(k)}$ and $q^{(k)}$ by
\begin{equation}\label{eq:udefine} 
	u^{(k)}=\frac{q^{(k)}}{\|q^{(k)}\|}\;, \quad \mbox{and} \quad
	q^{(k+1)}=A(\theta_0+k\om)u^{(k)}\;,
\end{equation}
This implies $\prod_{j=0}^{k} ||q^{(j)}|| = ||A^{(k)}(\theta_0) q^{(0)}||$, where $A^{(k)}$ is defined in \Eq{eq:Andef}. Since $\psi_1$ is the eigenfunction with the largest eigenvalue, we have
\begin{equation}\label{eq:uiteration}
    u^{(k)} = \frac{\lambda_1^k}{||A^{(k)}(\theta_0) q^{(0)}||}
                     \left[ \alpha_1 \psi_1(\theta_0+k\om) + 
                           {\cal O}\left(\frac{\lambda_2}{\lambda_1}\right)^k \right] \;,                 
\end{equation}
Because $u^{(k)}$ is a unit vector by definition and $\psi_1$ is continuous, its coefficient in \Eq{eq:uiteration} must be bounded. Defining $s_k = \sgn(\lambda)^{k}$, we therefore have
\[
    u^{(k)} \rightarrow s_k \frac{\psi_1(\theta_0+k\om)}{||\psi_1(\theta_0+k\om)||},
\]
which is the vector we were seeking (up to the choice in sign). Moreover, the Lyapunov multiplier is obtained simply from 
\[
     \mu_1 = \lim_{k \rightarrow \infty} \left(\prod_{j=0}^{k} ||q^{(j)}|| \right)^{1/k} \;;
\]
however, this process has shortcomings in certain applications that the following modifications help to alleviate.

To find $\psi_1(\theta_0)$ we select a subsequence $k_j$ on which $\|k_j\om\|_{\Z^r}$ approaches zero (using the pseudo-norm \Eq{eq:pseudonorm}). For the case of an invariant circle, $r=1$, the sequence $k_j$ can be constructed from the continued fraction expansion
\[
  \om = a_0 + 1/(a_1 + 1/(a_2 + \ldots)) = \left[a_0, a_1, a_2, \ldots\right] \;,
\]
with elements $a_i \in \Z_+$, for example, $\gamma^{-1} = \left[0, 1,1,1,\ldots\right]$. An optimal sequence $k_i$ corresponds to the denominators of the continued fraction convergents, and are given by the recursion relation
\begin{equation}\label{eq:cfsequence}
    k_{j+1}= a_j k_j + k_{j-1}\;, \quad j \in \Z_+ \;, \quad k_{0}= 1 \;, k_1 = 1\;.
\end{equation}
For $\gamma^{-1}$, this gives the Fibonacci sequence $\{ 1,1,2,3,5, \ldots\}$.
For higher dimensional cases, we could use the generalized Farey tree of \cite{Kim86}, which also generates the best approximants.

Given the sequence $k_j$, a polynomial interpolation on $u^{(k_j)}$ through the nodes $\theta_0 + k_j \om$ can be used to approximate the value of $\psi_1$ at $\theta_0$.  We stop the iteration once a desired accuracy is obtained. The Lyapunov multiplier $\mu_1$ can be rapidly and accurately computed by using  $\psi_1(\theta_0)$ in \Eq{eq:lyapunov}.

If the system has a negative dominant eigenvalue, then the direction of the iterates will oscillate, and for this reason, we normalize the $u^{(k_j)}$ so that one component is $+1$, say.  If interpolation is carried out in this way, and an approximation to $\psi_1(\theta_0)$ is accepted but future iterates of $\psi_1(\theta_0)$ oscillate in direction, then the eigenvalue must be negative. 

An alternative method for computing $\psi_1(\theta_0)$ is to begin with an arbitrary vector $q^{(0)}$ at a point $\theta_0-k\om$ and iterate it forward $k$ steps to $\theta_0$. However, this does not give an efficient method for estimating the error, and  if nothing is known about the system a priori, then it is difficult to select the appropriate $k$. Since, by \Eq{eq:uiteration},  the convergence rate of the iterates to $\psi_1$  is $\lambda_2/\lambda_1$, the error in the interpolant sequence $u^{(k_j)}$ should decrease geometrically whenever the Lyapunov spectrum is simple, and we use this as a convergence check.  Finally, if the interpolation scheme does not converge, then this can be a indication that the system is either  irreducible or has a dominant Lyapunov multiplier with multiplicity larger than one.  Naively accepting a given iterate $u^{(k)}$ as the approximation to $\psi(\theta_0+k\om)$ would then be a grave mistake.

Once we have obtained $\psi_1(\theta_0)$ and $\lambda_1$ to within a selected accuracy,  we generate $\psi_1(\theta)$ on a dense set of $\theta$ values by defining
\begin{equation}\label{eq:psiovertorus}
     \psi_1(\theta_0+k\om) \equiv    \lambda_1^{-k} A^{(k)}(\theta_0)\psi_1(\theta_0)
\end{equation}
The resulting function satisfies \Eq{eq:eproblem} since 
\begin{align*}
	A(\theta_0+k\om)\psi(\theta_0+k\om)
	    &= \lambda^{-k} A^{(k+1)}(\theta_0)\psi_1(\theta_0)\\
	    &= \lambda(\lambda^{-k-1} A^{(k+1)}(\theta_0)\psi_1(\theta_0)\\
	    &= \lambda \psi(\theta_0+(k+1)\om).
\end{align*}
When $A$ is reducible the function $\psi_1$ computed in this way should be continuous on the invariant torus. Under the assumption that the Lyapunov spectrum
is uniform, the associated eigenspaces are themselves continuous. This implies the continuity of the direction of $\psi_1(\theta)$, but its magnitude will only
be continuous when the function $c(\theta)$ defined by 
 \begin{equation}\label{eq:efunctdef}
		c(\theta_0+k\om) = \lambda_1^{-k} \|A^{(k)}(\theta_0) \psi_1(\theta_0)\| 
\end{equation}
is continuous. While $c(\theta)$ is bounded because $\|A^{(k)}(\theta)v(\theta) \| \sim \lambda^{k}$, its continuity is not completely obvious.  We will see,
however, that $c$ will be continuous in the examples presented in \Sec{sec:examples}. 

\subsection{Subdominant Eigenfunctions}\label{sec:hotelling}

The power method easily finds the dominant right eigenfunction and its corresponding eigenvalue. It is also easy to find the eigenfunction with the smallest multiplier by iterating with $(A(\theta))^{-1}$ and replacing $\om$ with $-\om$. Similarly, the dominant left eigenfunction $\phi_1(\theta)$ can be found using the power method for \Eq{eq:left}, or equivalently by using \Eq{eq:eproblem} upon replacing $A(\theta)$ with $A^*(\theta)$, and  $\om$ with $-\om$. However, to find intermediate eigenfunctions, another strategy must be used. Here we discuss two possibilities.

One such technique is a generalization of Hotelling's deflation method \cite{Ralston65}. In this method we modify $A(\theta)$ to create a new cocycle in which $\lambda_1$ has been replaced with zero.  The power method using this modified skew-product will then converge to $\left(\lambda_2,\psi_2(\theta)\right)$, the next largest eigenvalue/eigenfunction pair. The modified system can be constructed if we already know the dominant right and left eigenfunctions:  
\begin{lem} 
	If $A(\theta)$ is reducible and has dominant eigenvalue $\lambda_1$ with corresponding 
	right and left eigenfunctions  $\psi_1(\theta)$ and $\phi_1(\theta)$, 
	normalized such that $\phi_1^*(\theta)\psi_1(\theta)=1$, then the matrix
	\begin{equation}
		W_1(\theta)=A(\theta)-\lambda_1\psi_1(\theta+\om)\phi_1^*(\theta)\label{W}
	\end{equation}
	has the same eigenvalues and eigenfunctions as $A(\theta)$ except $\lambda_1$ has been replaced with
	$0$.
\end{lem}
\begin{proof} 
Since the left and right eigenfunctions corresponding to distinct eigenvalues are orthogonal by \Lem{lem:orthogonal}, the eigenfunctions of $W_1(\theta)$ are the same as those of $A(\theta)$. To see $W_1(\theta)$ and $A(\theta)$ share the same spectrum except for $\lambda_1$, apply $(C(\theta+\om))^{-1}$ on the left and $C(\theta)$ on the right of \Eq{W} arriving at
\[
	(C(\theta+\om))^{-1}W_1(\theta)C(\theta)=J-\lambda_1[e_1,0,\cdots,0]\;.
\]
\end{proof}
Consequently, the power method can be used with $W_1(\theta)$ to find $\psi_2(\theta)$.  Similarly, the corresponding left eigenfunction can be found using $W_1^*$. This method can be continued in principle to compute all of the eigenfunctions using successive deflations
\[
	W_k(\theta)=A(\theta)-\sum_{i=1}^k\lambda_i\psi_i(\theta+\om)\phi_i^*(\theta),\;k=2,\dots,n-1,
\]
thereby determining $\psi_3(\theta),\dots,\psi_n(\theta)$.

A second method to find subdominant eigenfunctions is iterative Gram-Schmidt orthogonalization. In lieu of making a rank-one modification of $A(\theta)$, this method removes any components in the direction of the dominant eigenfunction at each iterate of the power method.  Again, suppose that we have found the dominant right and left eigenfunctions. To determine $\psi_2$, begin as usual with a randomly chosen $q^{(0)}$ at an initial point $\theta$.  Since we assume $A$ is reducible, $q^{(0)}$ can be written in the eigenfunction basis as \Eq{eq:initialguess}.  Since $\psi_2$ must be orthogonal to $\phi_1$, we project out the $\psi_1$ component by applying the transformation
\begin{equation}\label{eq:ortho}
	q \mapsto q_{\perp} = q - 
	    \frac{\phi_1^*(\theta)q}{\phi_1^*(\theta)\psi_1(\theta) } \psi_1(\theta) \;.
\end{equation}
In principle, $q_\perp = \sum_{i=2}^n \alpha_i\psi_i(\theta)$ has no component in the $\psi_1$ direction and thus the power method applied to $q_\perp$ will converge to $\psi_2$. However, roundoff errors inevitably introduce a small component in the $\psi_1$ direction and so we reapply the orthogonalization after each iteration. Similarly $\phi_2$ can be found by projecting out the $\phi_1$ component and iterating with the adjoint matrix.

The orthogonalization technique to find the remaining subdominant eigenfunctions requires projecting out the previously-found dominant components.

\subsection{Stability Analysis}\label{sec:stabanal}

We first look at how errors in the computation of the dominant eigenfunctions affect  the deflation process.  Suppose that the dominant eigenfunctions have been obtained up to some error
\begin{align*}
	\hat{\psi}_1(\theta)& \equiv \psi_1(\theta)+\varepsilon\psi_2(\theta),\;\\
	\hat{\phi}_1(\theta)& \equiv \phi_1(\theta)+C\varepsilon\phi_2(\theta),
\end{align*}
where $C \in {\R}$. Here we keep only the terms in the subspaces spanned by the two most dominant eigenfunctions and also assume the errors to be of the same order of magnitude. 
Then the deflated matrix $W_1$ can be computed only approximately as
\[
	\hat{W}_1(\theta)\;=\;A(\theta)-\lambda_1\hat{\psi}_1(\theta+\om)
	\hat{\phi}_1^*(\theta) \;.
\]

Now suppose  $q_0=\alpha_1\psi_1(\theta)+\alpha_2\psi_2(\theta)$ is the initial guess in the computation of $\psi_2$.  The error at the first step is
\[
	\big{(}W_1(\theta)-\hat{W}_1(\theta)\big{)}q_0 =
	\lambda_1\varepsilon\big{(}\alpha_1\psi_2(\theta+\om)+C\alpha_2 \psi_1(	     
	     \theta+\om)\big{)} + {\cal O}(\varepsilon^2)={\cal O}(\lambda_1\epsilon)
\]
Continuing this analysis shows that the error increases by a factor of roughly $\lambda_1$ for each iteration.
Since the result at each step will be scaled by approximately $\lambda_2$, the error in the second eigenfunction due to the error in $\psi_1$ at the $j$th iterate will be ${\cal O} (\varepsilon \left(\lambda_1/ \lambda_2\right)^j)$.  This instability is also inherent in Hotelling's deflation method for constant matrices \cite{Ralston65}.

The deflation method can be modified to ameliorate this instability by applying the the projection \Eq{eq:ortho} to (approximately) remove the growing components in the $\psi_1$ direction. This process can be applied occasionally as a corrective procedure. For eigenfunctions corresponding to eigenvalues of decreasing magnitude, this corrective procedure must be applied increasingly often in order to maintain the desired accuracy. Of course, the correction \Eq{eq:ortho} can only be computed using approximations to the eigenfunctions.  Keeping this in mind, a straightforward calculation shows the error in the $(p+1)^{st}$ eigenfunction is of the same magnitude as the error made in the original $p$ dominant eigenfunctions.

Note that Gram-Schmidt orthogonalization amounts to applying \Eq{eq:ortho} at every iterate; therefore, Gram-Schmidt has similar stability properties to Hotelling's deflation with the projective refinement. So while the error in successive eigenfunctions for both methods is theoretically of the same order as the initial error in the dominant eigenfunctions, in practice the instabilities inherent in these two methods makes reliable computation of the subdominant eigenfunctions very difficult to achieve.  Indeed, in \cite{Ralston65} it is remarked that the analogous deflationary procedures for constant matrices can often only be employed effectively by experienced numerical analysts. The deficiencies of the deflationary methods have prompted us to seek more stable techniques that will appear in a forthcoming paper \cite{Wysham05}.
\section{Examples} \label{sec:examples}
As a first test of the power method, we construct a diffeomorphism on $\R^2 \times \T^r$ with a hyperbolic invariant torus whose manifolds are known explicitly. Using the coordinates $(x,y,\theta) \in \R^2 \times \T^r$, the map is defined by
\begin{equation}\label{eq:testmap}
	    f\left( x , y , \theta \right)=
	     \begin{pmatrix} 
	         1 -\delta +\delta x +\epsilon y 
	         \left( \lambda 	g(\theta + \om)-\delta g(\theta)\right)\\ 
	         \lambda y \\
	         \theta + \om 
	    \end{pmatrix} \;,
\end{equation}
where $g(\theta) = g(\theta+m)\; \forall m \in \Z^r$. The torus $\calT = \{(1,0,\theta): \theta \in \T^r  \}$  is invariant with rotation vector $\omega$.

Moreover, it is easy to see that the surfaces
\begin{align*}
   \calS &\equiv \{ y = 0\} \\
   \calU & \equiv \{ x = 1+\epsilon y g(\theta) \}
\end{align*}
are invariant under \Eq{eq:testmap}; if $\lambda >1 > \delta > 0$ these are the stable and unstable manifolds, respectively, of $\calT$.  The mapping is volume and orientation preserving when $\lambda = \delta^{-1}$.

The linearization of \Eq{eq:testmap} on the torus gives the matrix
\[
     A(\theta) = 
     \begin{pmatrix}\delta & \epsilon(\lambda g(\theta+\omega)-\delta g(\theta)) & 0 \\
          0 & \lambda & 0 \\
          0 & 0 & I 
     \end{pmatrix} \;
\]
For the case $r=1$, $\calT$ is an invariant circle, and the three $\omega$-unrelated eigenfunctions of  \Eq{eq:eproblem}, are
\begin{equation}\label{eq:testeigen}
    \psi_1 = \begin{pmatrix} \epsilon g(\theta) \\ 1 \\ 0 \end{pmatrix} \;, \;
    \psi_2 = \begin{pmatrix} 0 \\ 0 \\ 1 \end{pmatrix} \;,\;
    \psi_3 = \begin{pmatrix} 1  \\ 0 \\ 0 \end{pmatrix}\;,
\end{equation}
with the eigenvalues $\lambda$, $1$, and $\delta$,  respectively.

The power method of \Sec{sec:pwrmeth} works well to compute the eigenfunctions \Eq{eq:testeigen}. The convergence to $\psi_1$ is governed by \Eq{eq:uiteration}, which shows that the $k^{th}$ iterate of the approximate eigenfunction, $u^{(k)}$, is (up to normalization)
\begin{equation}\label{eq:testpwrits}
    u^{(k)} \propto \alpha_1 \psi_1 +\lambda^{-k}\alpha_2 \psi_2
	+\left(\frac{\delta}{\lambda}\right)^k\alpha_3 \psi_3. 
\end{equation}
Equations \Eq{eq:testeigen} and \Eq{eq:testpwrits} then imply that the error in the first component should decay like $1/ \lambda$ while the error in the third component should decrease as $\delta/ \lambda$. Figure \ref{fig:iterationerror} is a log-linear plot of the error in the first and third components of the $k^{th}$ approximation to the dominant eigenfunction.  As can be seen, the convergence rate is exactly as predicted.  Redoing the calculation with various values $\delta$ and $\lambda$ gives similar agreement with the theoretical prediction.
\InsertFig{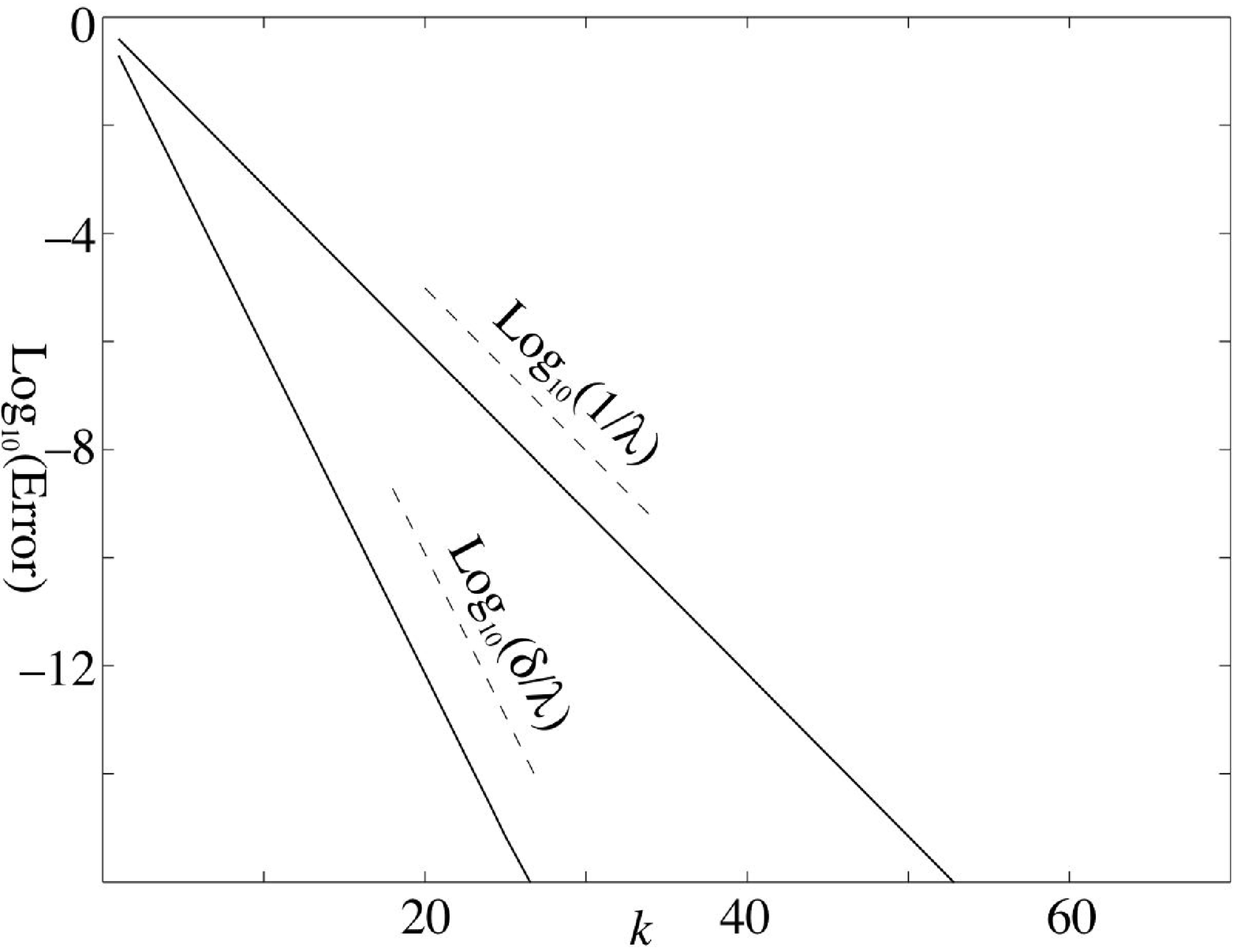}
	{Error in the first (upper curve) and third (lower curve) components of the eigenfunction 
	$\psi_1$ as a function  of $k$, the number of iterates, for the test system \Eq{eq:testmap}. Here 
	$g(\theta)= 0.7\cos \theta+ 0.13 \sin \theta $, $\lambda=1/\delta=2$, $\epsilon=.1$, 
	and $\om=\gamma^{-1}$.  Also shown are the predicted slopes of the errors.}
	{fig:iterationerror} {3in}

To compute $\psi_1(\theta_0)$, we interpolate as discussed in \Sec{sec:pwrmeth}, using continued fractions to generate the sequence \Eq{eq:cfsequence} of iterates $k_j$. Generally, the error in a $p^{th}$-order polynomial interpolation through nodes $x_0, \dots, x_p$ for a $C^{p+1}$ function at a point $x$ is
\[
	{\cal O}\left(\frac{1}{(p+1)!}\prod_{i=0}^{p}(x_i-x)\right) \;.
\]
We chose a third-order interpolation scheme. Since the sequence of best approximants to $\om$ obeys $|| k_j\om ||_\Z = {\cal O}(1/k_j)$, and the Fibonacci numbers grow as $\gamma^j$, the third-order interpolation error should scale as
\[
           {\cal O}\left(\gamma^{6-4j}\right).
\]
This is confirmed by our  computations, as shown in \Fig{fig:interpolationerr}. Note that the asymptotic slope is achieved only after the transient terms lying in the subdominant eigendirections have been eliminated (the second and third terms in \Eq{eq:testpwrits}). \Fig{fig:interpolationerr} demonstrates that more iterations are needed to remove this transient when the eigenvalue ratio $\lambda_1/\lambda_2 = \lambda$ is smaller. 

\InsertFig{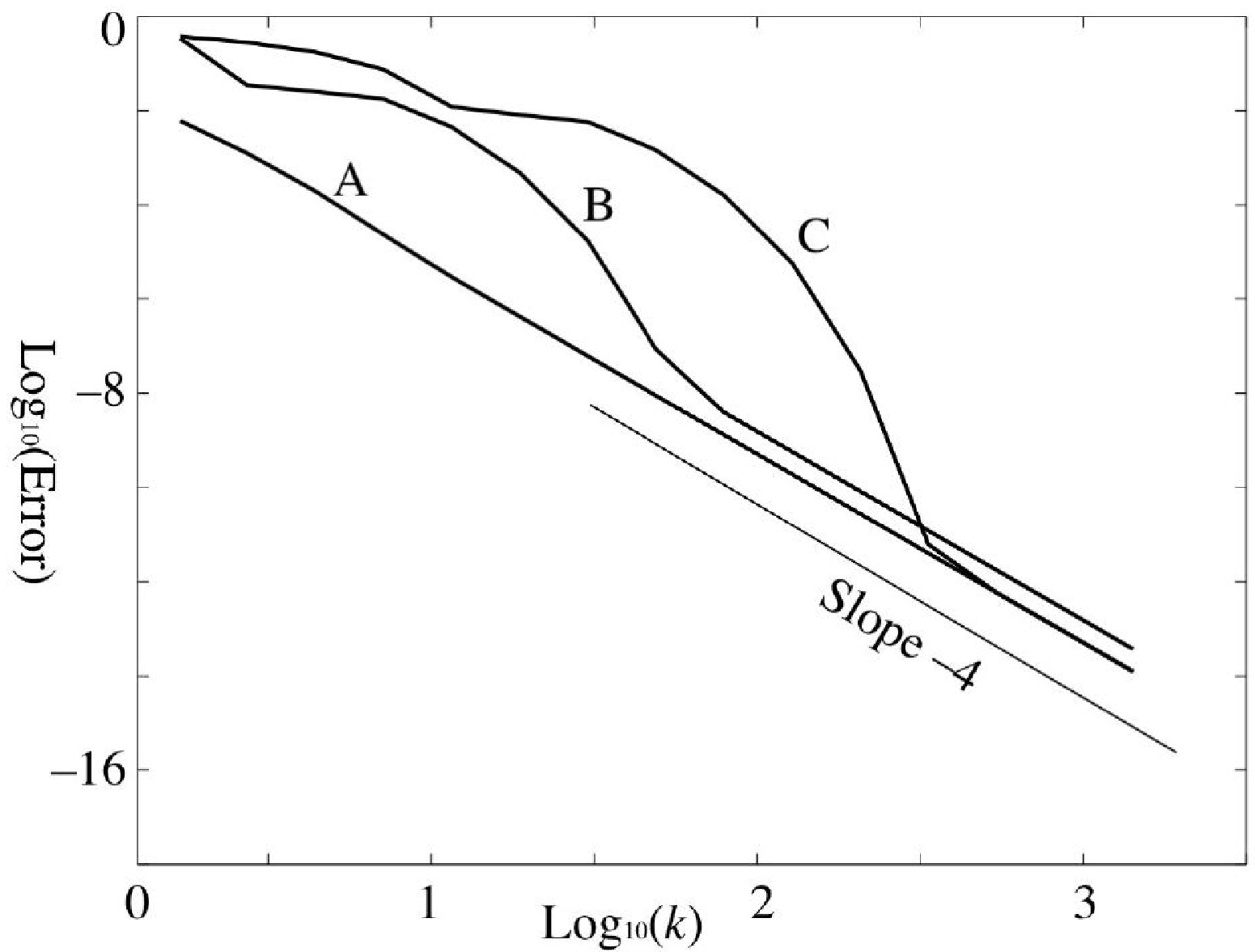}
	{Interpolation error in the computation of $\psi_1(0)$ for \Eq{eq:testmap}.  
	For these curves, $\lambda = \delta^{-1}$, $\om = \gamma$, and for curve
	(A) $\lambda = 2$, $\epsilon = 0.1$; (B)
	$\lambda = 1.5$, $\epsilon = 0.3$; and (C) $\lambda = 1.1$, $\epsilon = 0.1$. 
	The predicted slope of $-4$ is also shown for reference.}
	{fig:interpolationerr}{3in}

We accepted the value of $\psi_1(0)$ when successive interpolations varied by less than $10^{-16}$.  Subsequently, we set $\theta=0,$ and $v(0)=\psi_1(0)$ in \Eq{eq:lyapunov} in order to compute $\lambda_1$.  We accept the eigenvalue when the computations for the $k_i$ and $k_{i+1}$ iterations differ by less than $10^{-12}$.  The computed $\psi_1(0)$ and $\lambda_1$ were then used in \Eq{eq:psiovertorus} to generate $\psi_1(k\om)$ at $100$ locations round the invariant circle.  As shown in \Fig{fig:supthetaevecterr} the sup-norm of the final error in $\psi_1$ is less than $5\times 10 ^{-16}$. The sup-norm of the error in the eigenfunction $\psi_3$---computed using iteration with $A^{-1}$---is of the order $10^{-29}$.

\InsertFig{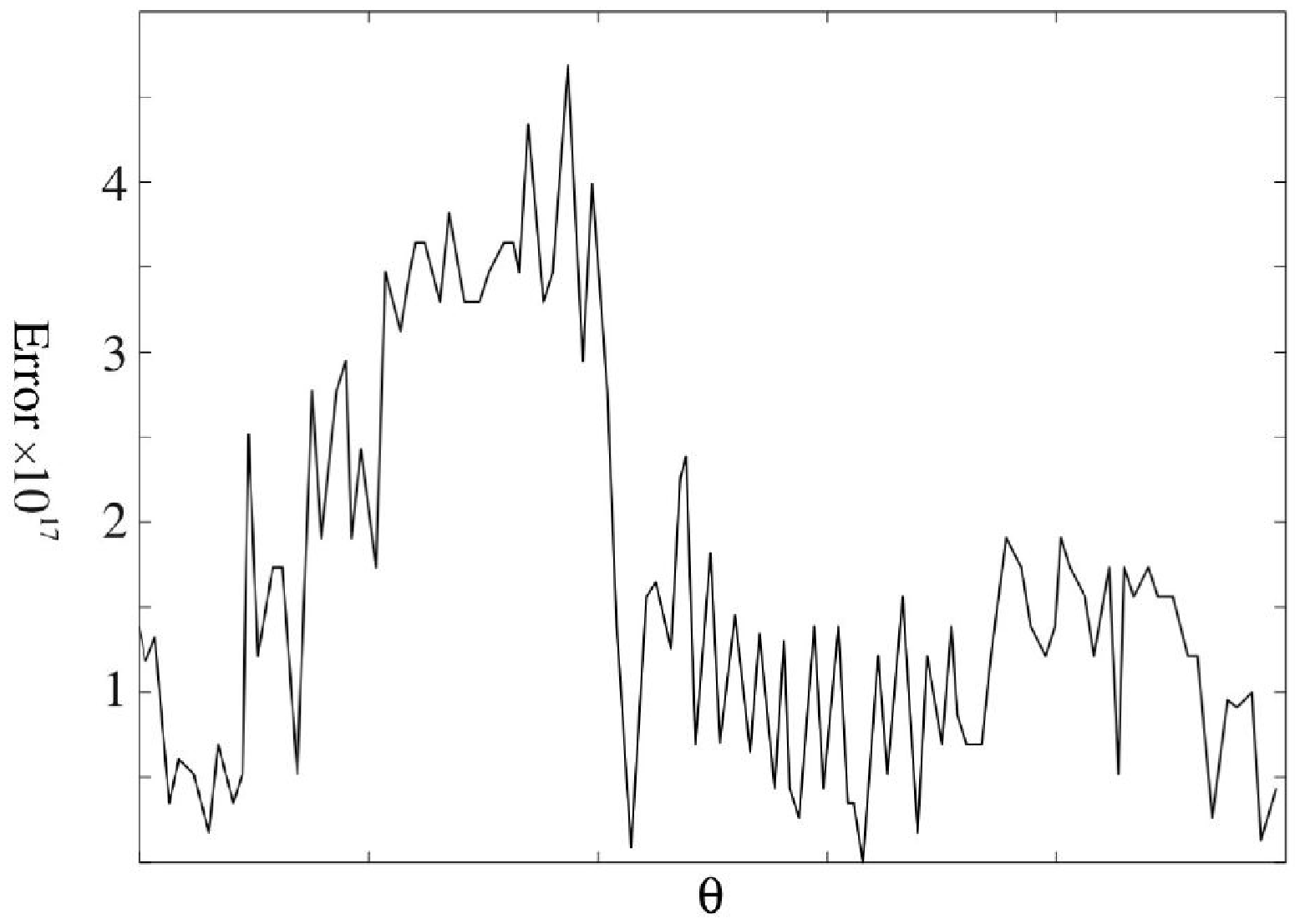}
	{Supnorm of the error in the numerical eigenvector $\psi_1(\theta)$ 
	for \Eq{eq:testmap} with the same  parameters as \Fig{fig:iterationerror}.}
	{fig:supthetaevecterr}{3in}

To test the method in a higher dimensional case, we set  $r=2$ in \Eq{eq:testmap}, so that the invariant torus is two-dimensional. The rotation vector was
taken to be $\om=(\tau,\tau^2)$, where $\tau$ is the real solution of the algebraic equation $x^3-x-1=0$.  This rotation vector is Diophantine since the
numbers $(1,\tau,\tau^2)$ form an integral basis for a cubic field \cite{Cusick72}. This time, instead of interpolating in order to find $\psi_1$ at a
particular value of $\theta$, we iterated $10^5$ times in order to remove any transient terms and simply accepted the last iterate as the approximation to $\psi_1$.  We recorded the location $\theta_0$ at which we have this approximation to $\psi_1$.  We then set $v(\theta_0)=\psi_1(\theta_0)$ in \Eq{eq:lyapunov} in order to compute $\lambda_1$, the value of which was accepted when successive approximations varied by less than $10^{-12}$.  Lastly, we generated $\psi_1(\theta_0+k\om)$ at $2\times10^4$ locations on the torus.  \Fig{fig:twotoruspsi} is a plot of the first component of the computed $\psi_1$.  Because the locations $\theta_0+k\om$ are irregularly spaced, Delaunay triangularization was used to compute the vertices of each triangular panel in the surface shown.

\InsertFig{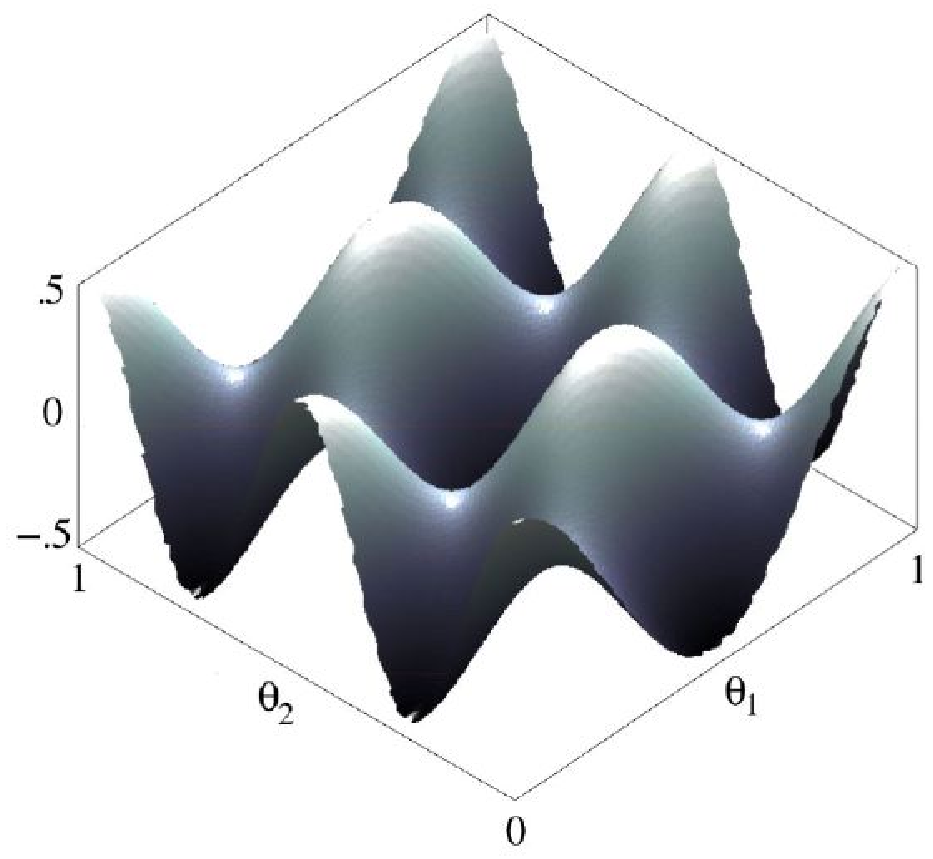}{Plot of the first component of $\psi_1(\theta_1,\theta_2)$ for the map \Eq{eq:testmap} with $\omega = (\tau, \tau^2)$. The choice of parameters was $\epsilon=.5,\lambda=1/\delta=1.5$, and $g(\theta)=\cos(2\pi\theta_1)\cos(4\pi\theta_2)$. The error in the computed values of $\psi_1$ was less than $5\times10^{-12}$.}{fig:twotoruspsi}{3in}

A more challenging test case is given by the volume-preserving example that appears in \cite{Lomeli00a, Lomeli03}. Here, the unperturbed diffeomorphism $f_0$ on $\R^3$ has a set of hyperbolic invariant circles whose invariant manifolds are also analytically known, but in this case they form heteroclinic connections between neighboring invariant circles. It was shown in \cite{Lomeli03} using a Melnikov method that this connection is destroyed upon perturbation and the topology of the resulting intersections of the invariant manifolds undergoes bifurcations as the parameters vary.  Our goal here is to explore the resulting manifolds numerically to verify and extend the predictions of the Melnikov theory. 

The unperturbed map $f_0$ in cylindrical coordinates $(r,\theta,z)$ is given by
 \begin{equation}\label{eq:lomeliMap}
	f_0(r,\theta,z) = \left(\begin{array}{c}
	 h^{-1}(r+h(z))-z \\ \theta+\omega \\ r+h(z)-1  \end{array} \right) \;,
\end{equation}
where the function $h(z)$  is an increasing circle diffeomorphism, i.e. $h(z+1)=h(z)+1\;\forall z \in {\R}$. For any such diffeomorphism, $f_0$ preserves the volume form $dr \wedge d\theta \wedge dz$. The map $f_0$ has an invariant circle at ${\calT} = \{ (1, \theta,z^*), \theta \in \bS\} $ where $z^*$ is a fixed point of $h$. The circle $\calT$ has invariant manifolds 
\begin{align*}
     \calU &= \{ r= h^{-1}(z)- h(z) + 1 \} \;, \\
     \calS &= \{ r = 1 \} \;.
\end{align*}
If $\lambda = \frac{1}{h'(z*)} > 1$, then $\calU$ is the unstable manifold of $\calT$ and  $\calS$ is its stable manifold. Moreover, if $h(z)$ has several fixed points, $z^{*}_i$, then $\calU$ and $\calS$ intersect at each of the corresponding invariant circles at $r=1$; thus, adjacent invariant circles have heteroclinic connections.  It is interesting that with a general choice of diffeomorphism $h$, the map $f_0$ appears to be nonintegrable even though it exhibits heteroclinic connections \cite{Lomeli03}.

Since  the cylindrical axis $r=0$ and the set $r>0$ are invariant under $f_0$, we can introduce rectangular coordinates $(x,y,z)$ using the volume preserving transformation
\begin{align*}
   (r,\theta,z) &\rightarrow (x,y,z) \;, \\
    x = \sqrt{2r} \cos(2\pi\theta) \;,\; y = &\sqrt{2r} \sin(2\pi\theta) \;,\; z = z \;,
\end{align*}
so that $r = \frac12 (x^2+y^2)$ is the ``symplectic" radius.
We use the same notation for the map in the new coordinates, letting $(x',y',z') = f_0(x,y,z)$. Now $f_0$ preserves the volume form $dx \wedge dy \wedge dz$.  

The linearization of $f_0(x,y,z)$ on the circle $\calT = \{ x^2+y^2 = 2, z=z^{*}\}$ gives rise to a skew product of the form \Eq{eq:skewprod} with the matrix
\begin{equation}\label{eq:LomeliA}
     A_0(\theta) = \begin{pmatrix} \cos(2\pi\om) + \delta \cos(2\pi\theta) \cos(2\pi(\theta+\om)) &
                                 -\sin(2\pi\om)+ \delta \sin(2\pi\theta) \cos(2\pi(\theta+\om))&
                                 0 \\
                                 \sin(2\pi\om) + \delta \cos(2\pi\theta) \sin(2\pi(\theta+\om))&
                                 \cos(2\pi\om)+ \delta \sin(2\pi\theta) \sin(2\pi(\theta+\om))&
                                 0 \\
                                 \sqrt{2} \cos(2\pi\theta) & \sqrt{2} \sin(2\pi\theta) &
                                  \lambda^{-1}\\
                \end{pmatrix}
\end{equation}
where $\delta = \lambda-1$.

It is straightforward to verify that eigenvectors of \Eq{eq:eproblem} for the matrix \Eq{eq:LomeliA} are
\[
      \psi_1 = \begin{pmatrix} \cos(2\pi\theta) \\ \sin(2\pi\theta) \\ 
              \sqrt{2} \frac{\lambda}{\lambda^2-1} \end{pmatrix} \;, \;
      \psi_2 = \begin{pmatrix} -\sin(2\pi\theta) \\ \cos(2\pi\theta) \\ 0 \end{pmatrix} \;,\;
      \psi_3 = \begin{pmatrix} 0\\0\\1 \end{pmatrix} \;,
\]
with eigenvalues $\lambda_1 = \lambda$, $\lambda_2 = 1$, and $\lambda_3 = \lambda^{-1}$. It is also easy to verify that $\psi_1$ is tangent to $\calU$, $\psi_2$ to the circle $\calT$, and $\psi_3$ to $\calS$.

An especially simple case occurs when $h$ is chosen to be the function \cite{Lomeli03}
\[
    h_{\nu}(z)=\frac{1}{\pi}\arctan\bigg{(}
       \frac{(\nu+1)\tan(\pi z)+\nu-1}{(\nu-1)\tan(\pi z)+\nu+1}\bigg{)} \;.
\]
In addition to being rotationally symmetric, $f_0$ now preserves the function
\begin{equation}\label{eq:invariant}
	 J(r,\theta,z) = 2 \nu \cos(2\pi r) + (1-\nu^2) \cos(2\pi z) \sin(2 \pi r) \;;
\end{equation}
consequently with this choice of $h$, $f_0$ is integrable.
In this case the surfaces $\calU$ and $\calS$ lie on the level set $J=2\nu$. This particular choice for the circle diffeomorphism is especially nice because iteration is easy: $h^{k}_{\nu}(z)=h_{\nu ^k}(z) \;,\; \forall k \in \Z$. The function $h_\nu$ has fixed points at $\pm z^*= \pm \frac{1}{4}$, with $h'(\pm z^{*}) = \nu^{\mp 1}$. When $\nu < 1$, the set $\calS$ is the stable manifold of the circle $\calT_-$ at $-z^*$ with eigenvalue $\nu$, and the unstable manifold of the circle $\calT_+$ at $+z^*$ with eigenvalue $\nu^{-1}$.

To destroy the heteroclinic connection between the circles  $\calT_{\pm}$, we perturb $f_0$ by composing it with a near-identity transformation, $Id + \epsilon P$. The resulting map is still volume preserving for all $\epsilon$ whenever the Jacobian $DP$ is nilpotent, i.e.,  $(DP_i)^3 \equiv 0$ \cite{Lomeli00a}.   We select two such transformations, so that
\begin{equation}\label{eq:perturbedLomeli}
        f_\epsilon= (Id+\epsilon P_1) \circ (Id + \epsilon P_2) \circ f_0 \;,
\end{equation}
where 
\begin{align*}
       P_1(x,y,z)  &= \left( (1+y^2)({z^{*}}^2 - z^2), 0,0\right) \;,\\
       P_2(x,y,z) &= \left( 0,0,r-1 \right) \;.
\end{align*}
Each of these $P_i$ clearly have nilpotent Jacobians, and they also vanish on the invariant circles $\calT_\pm$, so that no additional numerical work to find the circles is needed. Moreover, since the maps $Id+\epsilon P_i$ are diffeomorphisms, $f_\epsilon$ is a volume-preserving diffeomoprhism for all $\epsilon$. In this case the perturbed skew product has a matrix $A_\epsilon$ defined by
\[
    A_\epsilon(\theta) = \begin{pmatrix}
    				1-2z^*\epsilon^2 x'(1+(y')^2)  & -
				2z^*\epsilon^2y'(1+(y')^2) &-2z^*\epsilon(1+(y')^2) \\
				0 & 1 & 0 \\
				\epsilon x' & \epsilon y' & 1 
			\end{pmatrix}  A_0(\theta)
\]
where $(x',y') = (\sqrt{2} \cos(2\pi(\theta+\omega)), \sqrt{2}\sin(2\pi(\theta+\omega))$ corresponds to the rotated point on the circle.

To compute an eigenfunction, we begin with an arbitrary initial guess $q_0$.  Iteration is carried out using \Eq{eq:udefine}, normalizing the iterates to have sup-norm one at each iterate. As before we use the sequence  of convergents $k_i$ given by  \Eq{eq:cfsequence} and a third order polynomial interpolation to to find $\psi_1(0)$.  We continue iterating until the difference between successive interpolations differs by less than $10^{-12}$.

Once we find $\psi_1(0)$, we set $v(0)=\psi_1(0)$ in \Eq{eq:lyapunov} and compute the
corresponding eigenvalue.  Convergence to the eigenvalue is declared when computations at the ${k_i}^{th}$ and ${k_{i+1}}^{th}$ iterates vary by less than $10^{-8}$.  Once $\lambda$ is known, we then produce $\psi_1(\theta)$ at different locations on the torus using \Eq{eq:psiovertorus}. This creates a smooth eigenfunction in $\calT$, as can be seen in  \Fig{fig:efunctdef}; here we show $||\psi_1(\theta)||$ and  zoom-in near $\theta \approx 0.955$. A three-dimensional plot of $\psi_1$ constructed from 400 iterates of $\psi(0)$ for the same parameters is shown in \Fig{fig:psi1}. Similarly, backward iteration  with the power method can be used to find the stable eigenfunction. 

\InsertFig{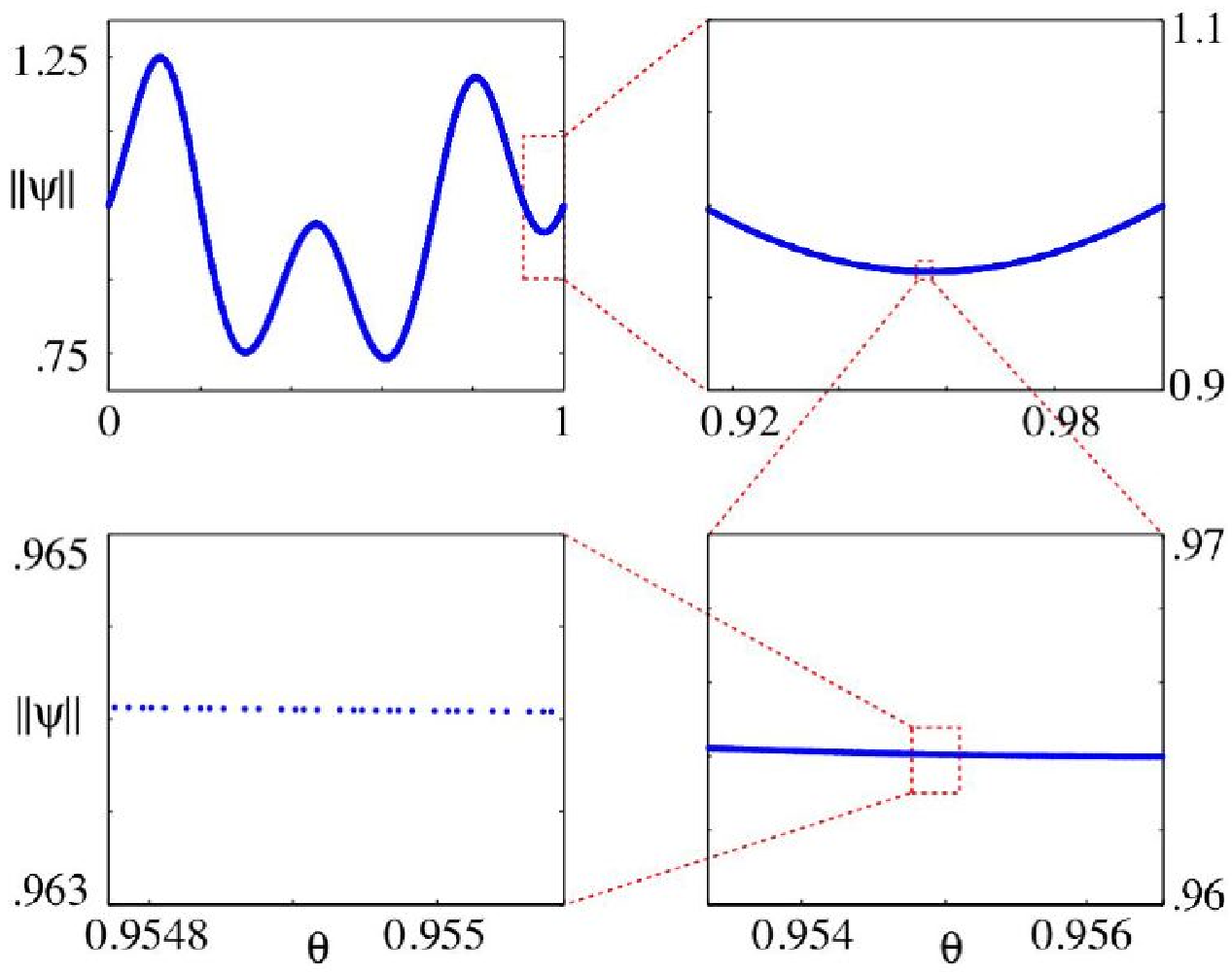}
	{The sup-norm of $\psi(\theta)$ for $\calT_-$ generated using \Eq{eq:efunctdef} for the map \Eq{eq:perturbedLomeli} with $\nu= 0.55$, $\omega=\gamma^{-2}$, and $\epsilon= 0.4$. To show that the generated eigenfunction is continuous, we zoom-in on a selected point. }
	{fig:efunctdef}{0.8\textwidth}

\InsertFig{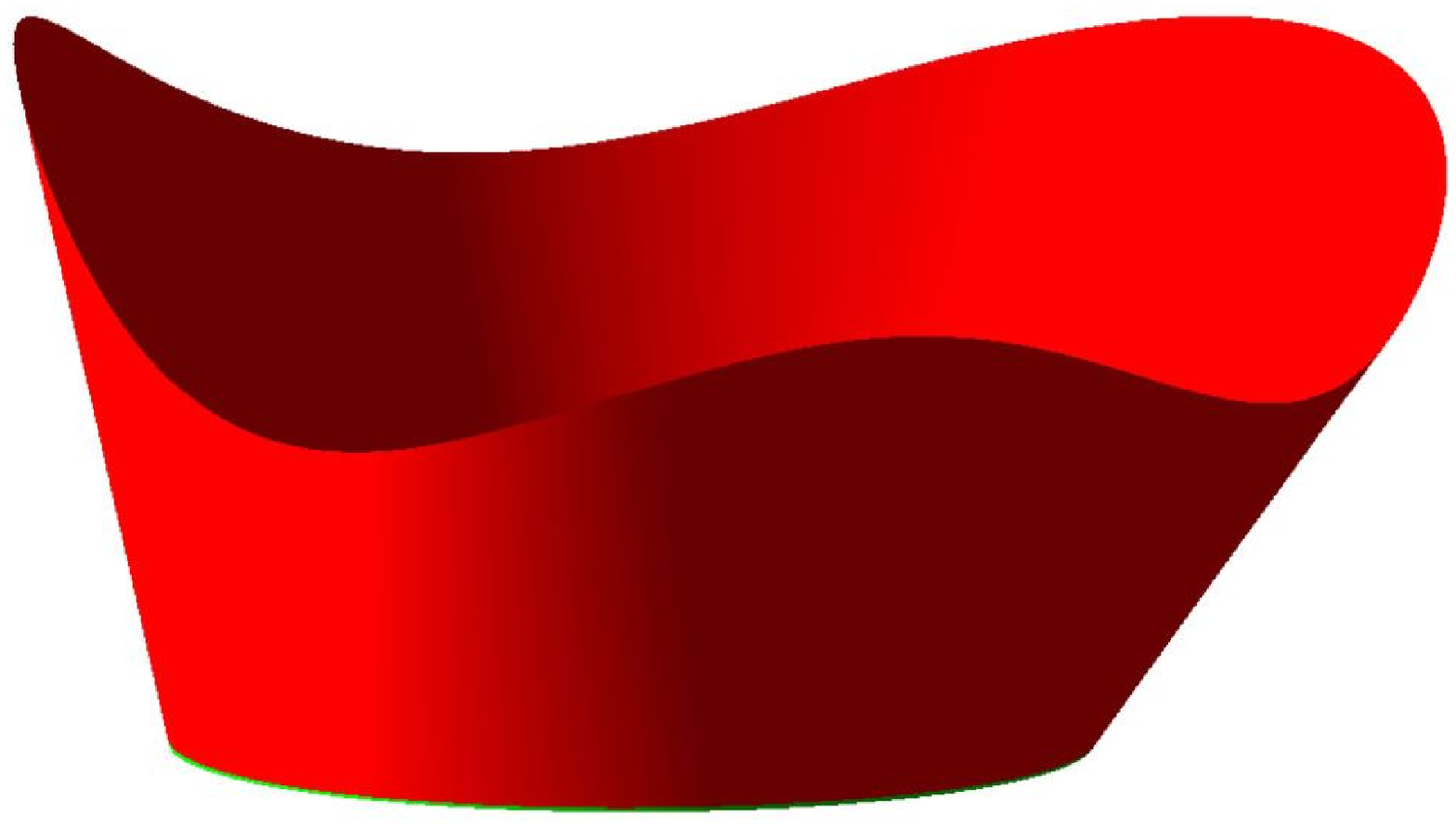}{Plot of the unstable eigenfunction $\psi_1$ on the circle $\calT_-$ for the map \Eq{eq:perturbedLomeli} with the same parameters as \Fig{fig:efunctdef}. }{fig:psi1}{0.8\textwidth} 

When $\epsilon = 0$ the eigenvalues are  $(\nu^{-1},1, \nu)$.  We show in \Fig{fig:lomeliEvalues} how they vary with $\epsilon$ for three different choices of
rotation number. Note that the eigenvalues appear to approach one at critical values of $\epsilon$. Since the perturbation is chosen to preserve the circle and
its rotation number, there is the slight possibility that these locations correspond to transcritical torus bifurcations. However, we have not been able to verify this as most
trajectories near the tori are unbounded.  Note that the power method becomes increasingly difficult to apply when the eigenvalues approach $1$ because the
linear convergence rate of the iterates $u^{(k)}$ is given by $\lambda_2/\lambda_1\approx 1$.  While this fact may explain some of the gaps in the graphs in
\Fig{fig:lomeliEvalues}, we are currently using more sophistocated techniques to determine exactly what happens dynamically in these regions.

\InsertFig{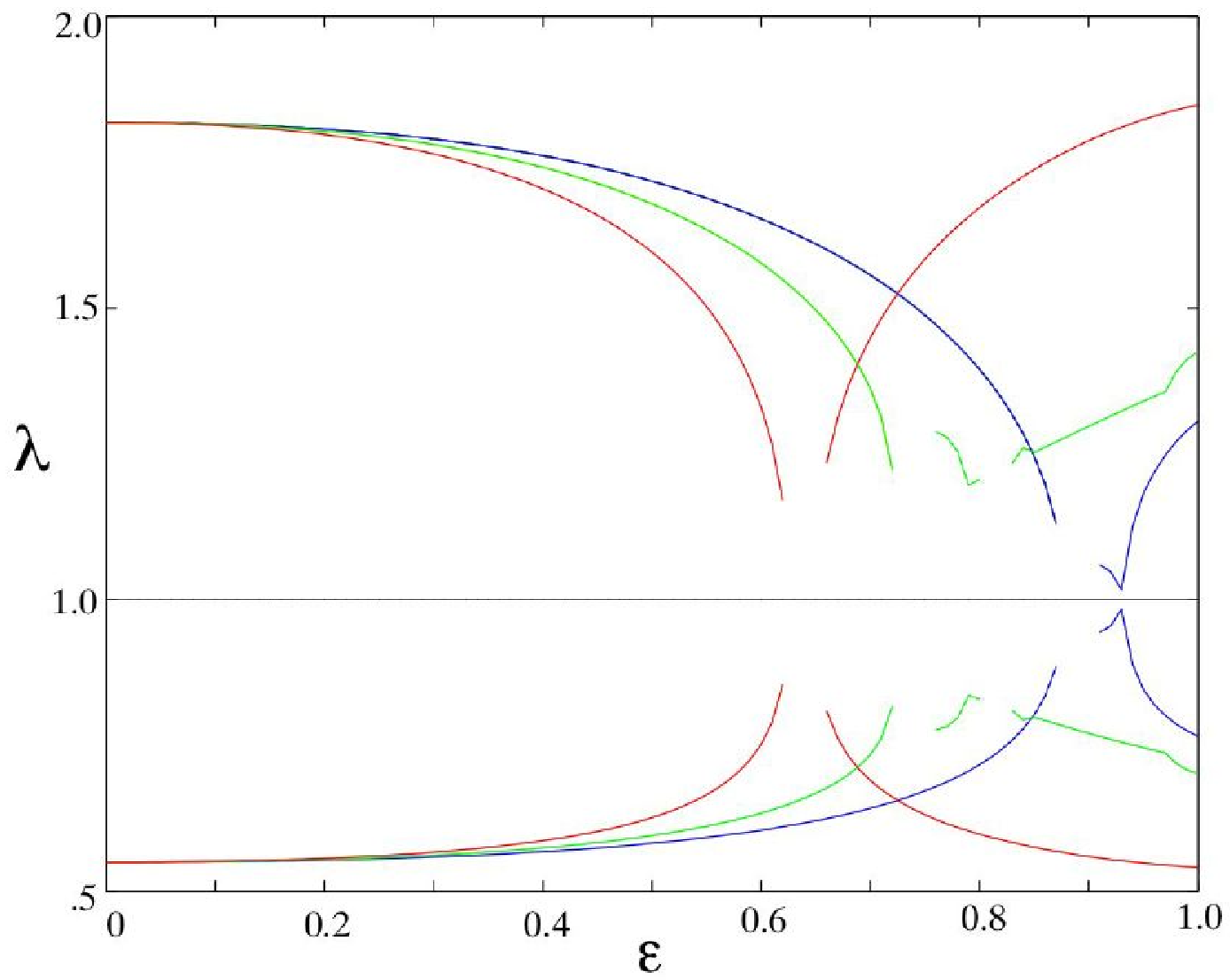}{Eigenvalues, $\lambda_i$, of the circle $\calT_-$ for the map \Eq{eq:perturbedLomeli} as a function of the perturbation strength $\epsilon$ with $\nu = 0.55$, and  $\omega = \gamma^{-2}$ (red curve), $\gamma^{-3}$ (green curve) and $\gamma^{-4}$ (blue curve)  as a function of $\epsilon$. For $\epsilon = 0$ $\lambda_1 = \lambda_3^{-1} = \nu$. Since the map is volume preserving the product of its eigenvalues is one.  }{fig:lomeliEvalues}{0.8\textwidth} 

Once we have obtained the eigenfunctions $\psi_i(\theta)$, we can plot the manifolds iteratively. Here we will compute the unstable manifold of $\calT_+$ and the stable manifold of $\calT_-$. To draw the manifolds we first construct an approximate {\em fundamental domain} \cite{Lomeli00a}.  A fundamental domain for a two-dimensional invariant manifold is an annulus $\calS$ with the property that the components, $c_i$, of its boundary, $\partial\calS = c_1\cup c_2$, are related by iteration, $f(c_1)=c_2$.  Using this fact, a fundamental domain can be iterated forward or backward to form an entire branch of the two-dimensional unstable or stable manifold of an invariant set.

For the case of an unstable manifold, we form an approximate fundamental domain from $\psi_1$ by choosing a factor $\rho \ll 1$ and defining a mesh of points from $\rho\lambda^{k/m}\psi(\theta_j),\;k=0,\dots,m$, $j = 0,\dots,N$ so that the inner loop, $c_1$, corresponds to the points with $k=0$ and the outer to $k=m$. Quadrilateral panels are drawn using OpenGL to connect the points on this mesh, and then each point on this mesh is iterated forward to form a new annulus $f(\calS)$ that is concatenated onto $\calS$. Panels are again drawn connecting the new grid of points on $f(\calS)$ and the process is repeated. In our computations we choose $\rho = 10^{-6}$, $m = 5$, and $N=400$.

In the mapping investigated here, the manifolds have transversal intersections as was predicted by the Melnikov calculations in \cite{Lomeli03}.  For this reason, we cannot iterate a fundamental domain too many times before the oscillations of the manifolds become wild and the pictures too difficult to interpret. The number of iterates will be of order $-\log \rho / \log \lambda$ for the manifolds to grow to a size of order 1. We used $14$ iterates in the plots shown in \Fig{fig:manifolds}(a) and (b), $12$ for (c), and $27$ for (d).

\InsertFigFour{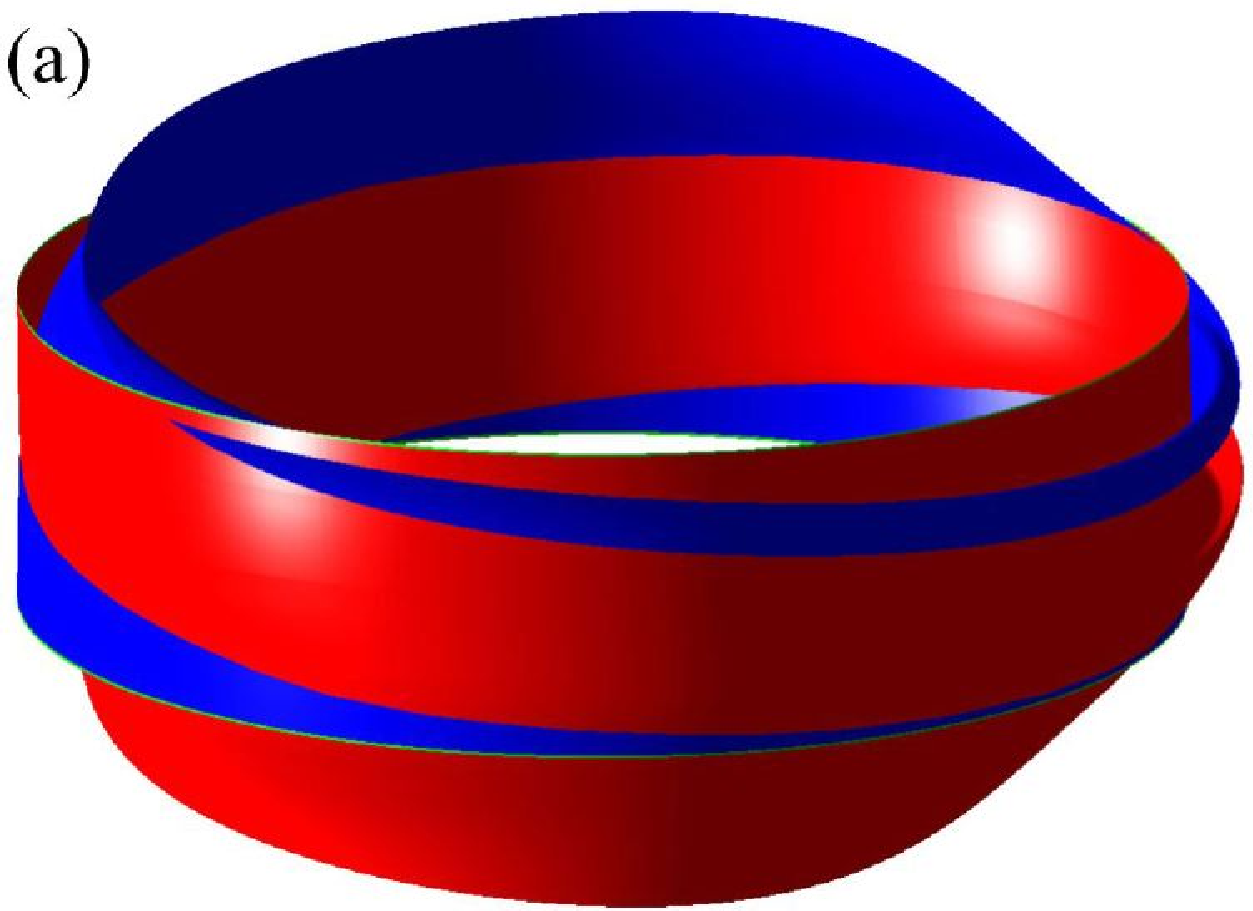}{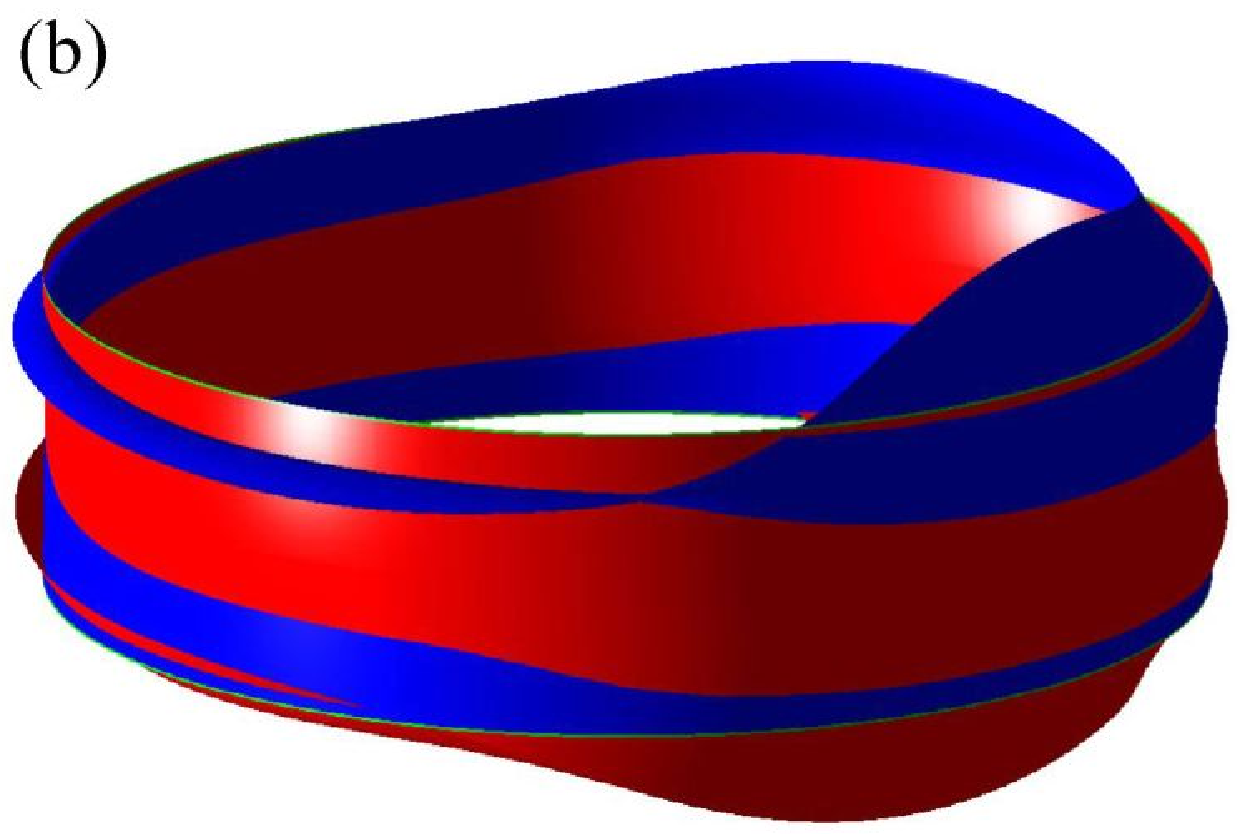}{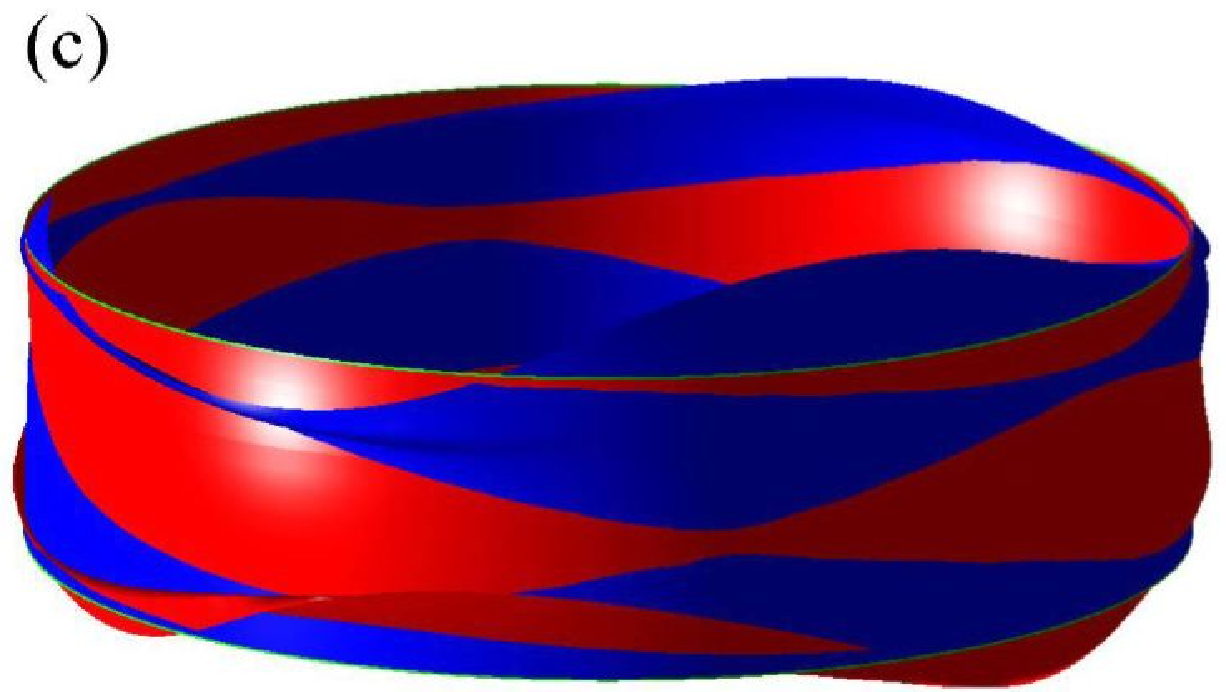}{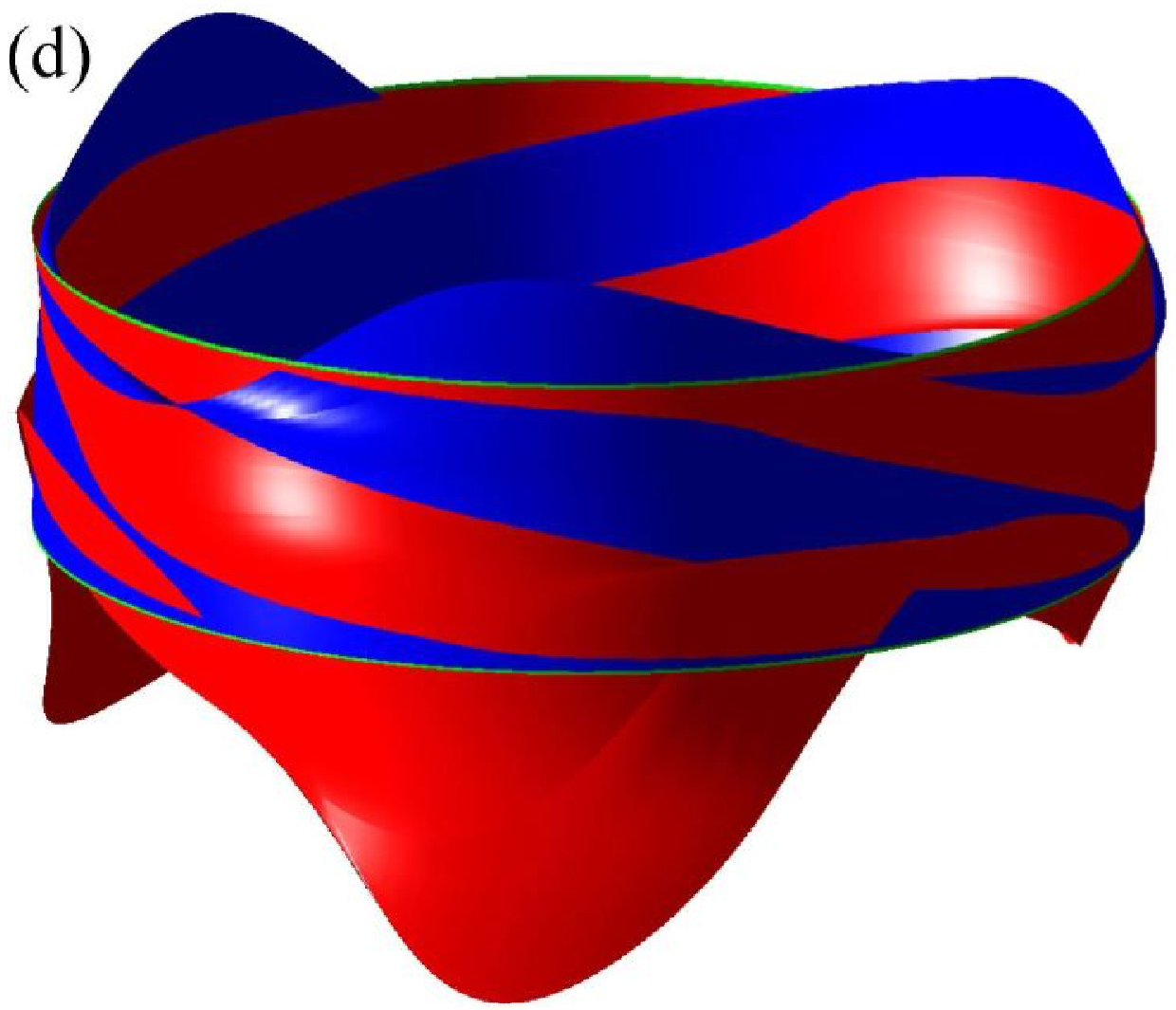}
{Stable manifold (blue) of the circle $\calT_-$ and unstable manifold (red) of the circle $\calT_+$ for the map \Eq{eq:perturbedLomeli}. The top two panels have parameters  $\nu = 0.35$ and  $\epsilon = 0.25$ with (a) $\omega = \gamma^{-4}$, and  (b)  $\omega=\gamma^{-3}$. The bottom two panels have $\omega = \gamma^{-2}$ with (c) $\nu = 0.30$ and $\epsilon = 0.35$, and (d) $\nu = 0.55$, and $\epsilon = 0.4$.}{fig:manifolds}{3in}

In \Fig{fig:manifolds} we show the invariant circles $\calT_{\pm}$ (green curves) as well as the unstable manifold of $\calT_+$ (red surface) and the stable manifold of $\calT_-$ (blue surface). The first two panels, (a) and (b), show  the manifolds intersecting along an infinite spiral that is forward asymptotic to $\calT_-$ and backward asymptotic to $\calT_+$. In the last two panels, (c) and (d) the intersection curves undergo a reconnection.

Since the images of each fundamental annulus $\calS$ cover the invariant manifold, if there are intersection curves, they must go through each fundamental annulus on each manifold.  A fundamental annulus can be considered topologically to be a torus if we identify the boundary $c_1$ with $c_2$ using  the fact that they are images under $f$.  Thus we can classify the intersection curves as elements of the homology group, $\Z^2$, of the torus. The intersection curves in \Fig{fig:manifolds}(a) and (b) have homology $(1,0)$, crossing the fundamental domain once vertically with no net rotations in $\theta$. As $\omega$ is increased to values near the resonance $1/3$ there is a homology bifurcation. This is shown in \Fig{fig:manifolds}{c} where  $\omega = \gamma^{-2} \approx 0.382$. Though the intersection curves still have $(1,0)$ homology in this case, they are nearly touching. A reconnection does occur in \Fig{fig:manifolds}(d) as $\nu$ is increased, causing the homology to become $(3,1)$, corresponding to a three-armed spiral.

The intersections of the stable and unstable manifolds for $f_\epsilon$ when $\epsilon \ll 1$ can be studied using the Melnikov technique \cite{Lomeli03}. The Melnikov function, $M: W_0^{u}(\calT_+) \rightarrow \R$, is measure of the distance between the stable, $W_\epsilon^s(\calT_-)$ and unstable $W_\epsilon^u(\calT_+)$ manifolds to first order in $\epsilon$. As is shown in \cite{Lomeli03}, the Melnikov function is given by
\begin{equation}\label{eq:melnikov}
      M = \sum_{t=-\infty}^{\infty} dJ(X_0) \circ f_0^t \;.
\end{equation} 
Here, $dJ$ is the one-form defined by the invariant $J$, \Eq{eq:invariant}, and $X_0$ is the ``perturbation vector field"
\begin{align*}
    X_0(x) \equiv \left. \frac{\partial}{\partial \epsilon} f_\epsilon(f^{-1}_0(x)) 
           \right|_{\epsilon = 0}
           = P_1 + P_2\;.
\end{align*}
If $M$ has a nondegenerate zero crossing then for $\epsilon$ small enough there is a true transversal intersection of $W^u_\epsilon$ and $W^s_\epsilon$ nearby. We plot contours of $M$ in \Fig{fig:melnikov} for the same set of parameters as \Fig{fig:manifolds}. In this figure the domain corresponds to the fundamental domain on $W_0^u$, given by $\{0 \ge z \ge h_\nu(0) \;,\; 0 \le \theta < 1\}$, and the  zero level set of $M$ corresponds to the heavy curves. The identification of the upper boundary with the lower boundary to form a torus is done by shifting these boundaries by $\omega$, as indicated by the arrows. Note that the homology types of the zero level sets are the same as those observed in \Fig{fig:manifolds}.

\InsertFigFour{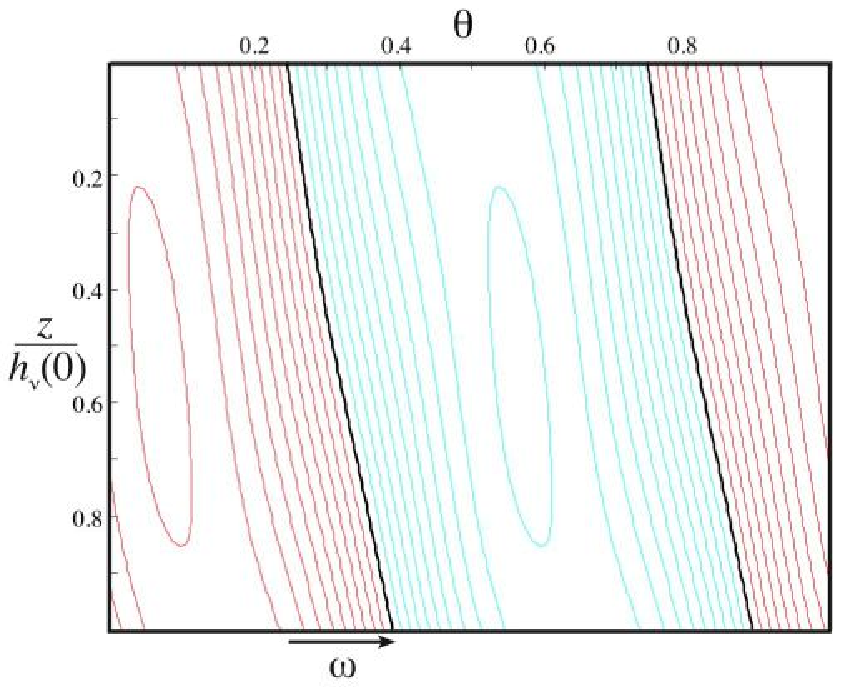}{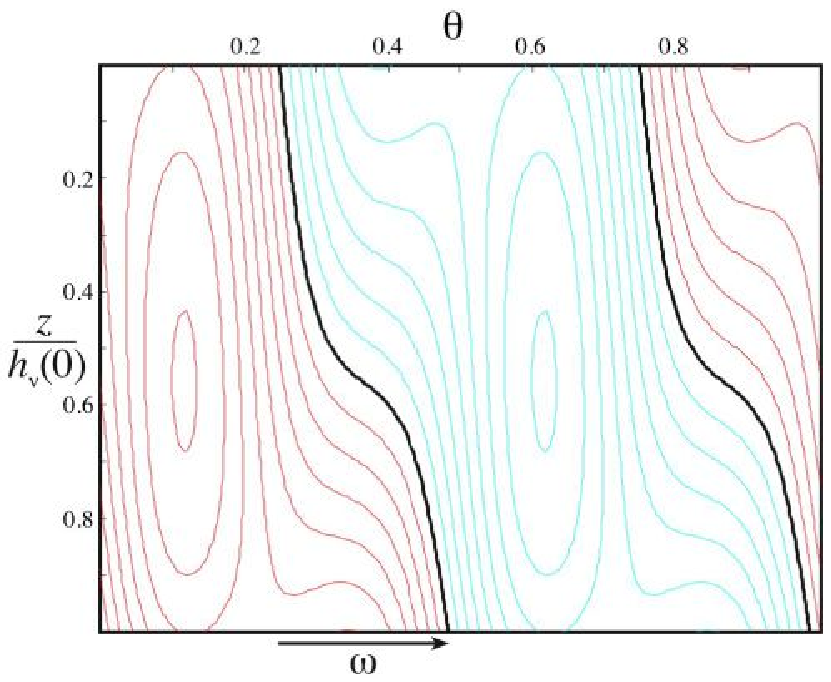}{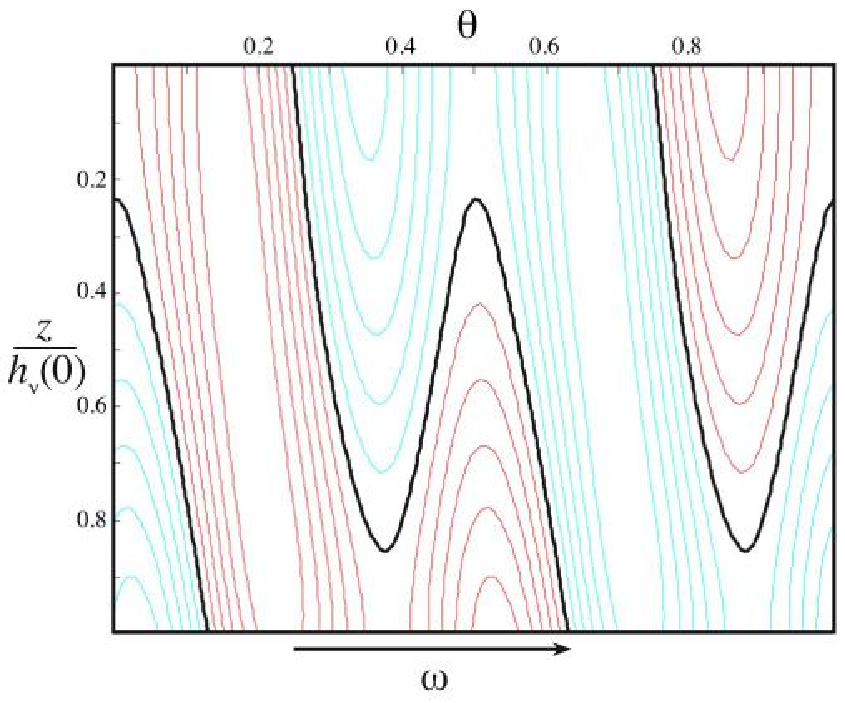}{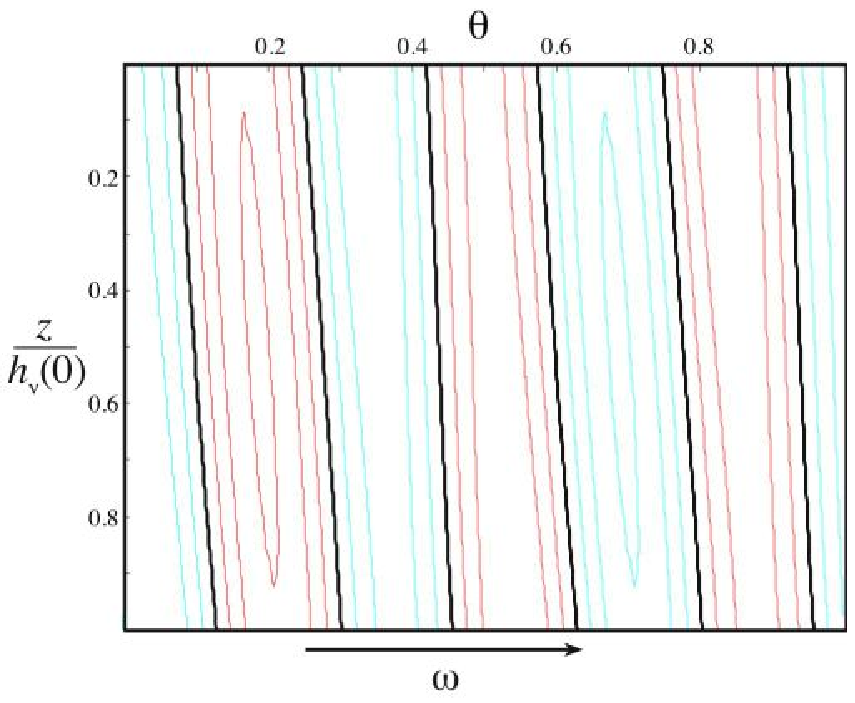}{ Contour plots of the Melnikov function \Eq{eq:melnikov} on a fundamental domain of $W_0^u(\calT_+)$  for the same set of parameter values as \Fig{fig:manifolds}. In these figures the zero level set is the heavy (black) curve, while positive and negative levels of $M$ are colored red and cyan, respectively. }{fig:melnikov}{3in}

\section{Conclusion}
The linearization of a mapping near an invariant torus with an incommensurate rotation vector gives rise to a quasiperiodic skew-product of the form \Eq{eq:skewprod} with a periodic matrix $A(\theta)$. If one can find a coordinate transformation in which the skew-product is rewritten as a constant matrix, then the system is called reducible. The eigenvalues and invariant directions of the $A(\theta)$ correspond to the linear invariant subspaces of the torus. Because these provide the first order approximations to the invariant manifolds for the invariant torus, these methods can be used in computing the manifolds of reducible invariant tori. We discussed the eigenvalue problem \Eq{eq:eproblem} associated with the reducing transformation, and showed how to define generalized eigenfunctions to reduce the system when it has multiple eigenvalues.

We applied our methods to compute the invariant manifolds for several examples, including a system studied in \cite{Lomeli03} by Melnikov methods. Our computations confirm the bifurcations in homology of the heteroclinic intersections between a pair of invariant circles predicted by \cite{Lomeli03}.

While Fourier series methods have been previously used in numerical algorithms to compute the eigenfunctions, our simple iterative technique based on the power method works well when the eigenspaces are one dimensional. Intermediate eigenfunctions can also be found using iterative deflation methods, though their numerical stability properties---inherited from the constant matrix case---make them problematic. In the future we will present methods based on an iterative $QR$ decomposition which much more effectively deflate $A(\theta)$. Algorithms for the cases of complex and multiple eigenvalues will also appear in that work.

\bibliographystyle{alpha}

\end{document}